\documentclass[%
 reprint, 
 superscriptaddress,
preprintnumbers,
 amsmath,amssymb,
 aps, 
]{revtex4-2}
\usepackage{graphicx}% Include figure files
\usepackage{dcolumn}% Align table columns on decimal point
\usepackage{bm}% bold math
\usepackage{color}

\begin{document}

%\preprint{}

\title{Dynamically Polarized SERF Atomic Comagnetometer}



% \author{Xiaofei Huang,$^{1,2}$ Kai Wei,$^{1,2,3,*}$ Yang Rui,$^{4}$ Saixin Zhou,$^{1,2}$ Jie Zheng,$^{1,2}$ and Wei Quan$^{1,2,3,*}$}
% \affiliation{$^1$School of Instrumentation Science and Opto-electronics Engineering, Beihang University, Beijing 100191, China}
% \affiliation{$^2$Hangzhou Innovation Institute, Beihang University, Hangzhou 310051, China}
% \affiliation{$^3$Hefei National Laboratory, Hefei 230088, China}
% \affiliation{$^4$Hangzhou Extremely Weak Magnetic Field Major Science and Technology Infrastructure Research Institute, Hangzhou, 310051, China}

\author{Xiaofei Huang}
\affiliation{School of Instrumentation and Optoelectronic Engineering, Beihang University, Beijing 100191, China}
%\affiliation{Hangzhou Innovation Institute, Beihang University, Hangzhou 310051, China}

\author{Kai Wei}
\email[]{Contact author: weikai@buaa.edu.cn}
\affiliation{School of Instrumentation and Optoelectronic Engineering, Beihang University, Beijing 100191, China}
%\affiliation{Hangzhou Innovation Institute, Beihang University, Hangzhou 310051, China}
\affiliation{Hefei National Laboratory, Hefei, 230088, China}

\author{Yang Rui}
\affiliation{Hangzhou Institute of Extremely-Weak Magnetic Field Major National Science and Technology Infrastructure, Hangzhou, 310051, China}

\author{Dinghui Gong}
\affiliation{School of Instrumentation and Optoelectronic Engineering, Beihang University, Beijing 100191, China}

\author{Saixin Zhou}
\affiliation{School of Instrumentation and Optoelectronic Engineering, Beihang University, Beijing 100191, China}
%\affiliation{Hangzhou Innovation Institute, Beihang University, Hangzhou 310051, China}

\author{Jie Zheng}
\affiliation{School of Instrumentation and Optoelectronic Engineering, Beihang University, Beijing 100191, China}
%\affiliation{Hangzhou Innovation Institute, Beihang University, Hangzhou 310051, China}

\author{Wei Quan}
%\email[]{Contact author: quanwei@buaa.edu.cn}
\affiliation{School of Instrumentation and Optoelectronic Engineering, Beihang University, Beijing 100191, China}
%\affiliation{Hangzhou Innovation Institute, Beihang University, Hangzhou 310051, China}
\affiliation{Hefei National Laboratory, Hefei, 230088, China}


\date{\today}

\begin{abstract}

\textbf{Summary:}

Atomic spin sensors are essential for beyond-the-standard-model exploration, biomagnetic measurement, and quantum navigation. 
While the traditional DC mode spin-exchange relaxation-free (SERF) comagnetometer achieves ultrahigh sensitivity, further improvements require suppressing technical noise and surpassing standard quantum limit.
In this work, we develop a K-Rb-$^{21}$Ne SERF atomic comagnetometer that dynamically polarizes the electron and nuclear spins, shielding signals from direct interference by pump light. 
We establish a three-phase evolutionary model for hybrid spin ensemble dynamics, yielding a complete analytical solution, and  analyze the responses to various spin perturbations.
Additionally, we achieve an averaged 38.5 $\%$ suppression of the polarization noise and identify the key factors that limit sensitivity improvements. 
The dynamically polarized comagnetometer exhibits effective suppression of technical noise and holds the potential to overcome quantum noise limit, while offering promising applications in exploring new physics and precise magnetic field measurements.\\


\end{abstract}


\maketitle

\section{Introduction}
\label{Sec:Intro}
% \subsection{\uppercase\expandafter{\romannumeral1}~Introduction}
%\subsubsection*{}

In alkali-noble-gas sensors, collocated alkali electron spins and noble gas nuclear spins are coupled via spin-exchange interactions~\cite{Wei2023Ultrasensitive}. This combination of spin species with distinct relaxation and precession characteristics facilitates the development of self-compensation comagnetometers~\cite{ji2018new,Klinger2023Optimization,Almasi2020New}, clock-comparison comagnetometers~\cite{Feng2022Search, Zhang2023Search}, and hybrid-spin-resonance comagnetometers~\cite{Wei2025Dark,heng2025search}. These comagnetometers exhibit ultrahigh sensitivity across different frequency bands, enabling the exploration of ultralight axion-like particles and Fifth force~\cite{Safronova2018, Terrano2021, JacksonK2023, huang2024axionlike}, quantum navigation~\cite{zhang2020closed,fang2012advances}, and magnetic field measurement~\cite{MARSHALL202142,jiang2021search}.

Light-atom interactions provide the necessary spin polarization and detection, which is essential for the operation of various quantum sensors~\cite{walker1997spin}. However, in traditional DC mode, the instability of pump light leads to undesirable signal variations ~\cite{Sato2018}, and the photon fluctuations in the probe light give rise to the standard quantum limit~\cite{Matsuzaki2011Magnetic}. These significantly hinder the improvement of sensitivity. Dynamic nuclear polarization via electron-to-nuclear polarization transfer can enhance nuclear spin polarization, thereby improving the sensitivity and spatial resolution of nuclear magnetic resonance experiments \cite{can2015mechanisms, akbey2016structural}. Light pulses and microwave fields further enhance nuclear spin polarization in nitrogen-vacancy center in diamond\cite{xu2019dynamically, ajoy2018enhanced}. Ref. \cite{smirnov2020dynamic} proposes a dynamic electron spin polarization method in semiconductor nanostructures driven by nuclear spin fluctuations. The stroboscopic quantum non-demolition measurements demonstrate the capability for dynamic manipulation that exceeds the standard quantum limit~\cite{bao2020spin, vasilakis2015generation}.
Thus, the dynamic polarization process resulting from interactions between electron spins and nuclear spins demonstrates significant importance in quantum communication and quantum sensing. Dynamic polarization schemes based on pulsed pump light have garnered significant attention in the fields of atomic magnetometers and atomic clocks~\cite{borna2018magnetic, arditi1964atomic, hao2019microwave}. The spin precession that follows the turning off of the pump light is advantageous for sensing magnetic fields, as the precession frequency is directly related to the magnetic field~\cite{hopf1973theory, Hunter2022}. Light-induced noise and light shifts are effectively suppressed during the probe phase due to the absence of pump light~\cite{jaufenthaler2021pulsed, kang2015demonstration}.

In spin-exchange relaxation-free (SERF) atomic comagnetometers, polarization noise is regarded as the primary low-frequency noise source, thereby impeding the improvement of sensitivity~\cite{Xu2024Analysis}.
Recently, research on a perturbed $^{87}$Rb-$^{129}$Xe comagnetometer scheme in SERF regime utilizing synchronous pump and magnetic pulses demonstrated the feasibility of dual-axis inertial rotation and magnetic field measurements based on the free precession of alkali electron spins~\cite{Hedges2025Dual}. Ref.~\cite{wang2025pulsed} presented a $^{87}$Rb-$^{21}$Ne comagnetometer using pulsed optical pumping, achieving dual-axis measurements and suppression of pump and probe beam pointing fluctuations, while emphasizing that improving nuclear spin relaxation, which is currently limited by the gradient of dipolar field, is key to enhancing sensitivity. 
Both of these studies employ a single species of alkali metal, which simplifies the analysis of electron spins evolution. However, due to the high atomic density required in the SERF regime, achieving homogeneous polarization of electron spins necessitates extremely high intensity to ensure complete polarization throughout the entire cell. Under such high pump light intensity, the heating effect of the pump light~\cite{Xu2021Fast} induces temperature fluctuations in the vapor cell during the switching process, thereby introducing additional perturbations.
Furthermore, the evolution of dynamically polarized comagnetometer based on pulsed pump light is complex, and a thorough mechanism analysis is crucial for improving the performance.
A common assumption in traditional DC mode is that the longitudinal components of spin species are approximately independent of the transverse components~\cite{kornack2005nuclear}. Unfortunately, this approximation does not hold in our dynamic polarization scheme because the polarization of alkali electron spins varies drastically due to the switching of the pump light. Moreover, the precession frequency shifts with changes in the Zeeman sub-level populations. Although the pulsed pump light introduces complexities in dynamic behavior, the coupling phenomena of spin ensemble are nonetheless intriguing and warrant further investigation.

In this work, we develop a K-Rb-$^{21}$Ne dynamically polarized SERF comagnetometer utilizing pulsed pump light. 
The hybrid pumping we utilized can improve the homogeneity of alkali electron spins, achieving the reduction of the electron effective magnetic field gradient and, consequently, enhancing the nuclear spin relaxation time~\cite{Wei2023Ultrasensitive}. Based on the distinct relaxation and resonance frequencies of alkali electron spins and noble gas nuclear spins, we establish a three-phase evolutionary model that enables us to derive a comprehensive analytical solution for the entire dynamic polarization process. Leveraging this model, we explore a method for magnetic compensation under dynamic polarization. Subsequently, we analyze the responses to dual-axis inertial rotation and low-frequency magnetic fields at the compensation point and experimentally validate these responses at five duty ratios. By increasing the duty ratio from 40 \% to 60 \%, we achieve an enhancement in the longitudinal polarization of nuclear spins, resulting in an average suppression of the low-frequency magnetic field response by 51.1 \%. Furthermore, we examine the suppression of polarization noise via pump light intensity modulation, which yields an averaged suppression of 38.5 $\%$, and investigate the limiting factors that affect sensitivity under the dynamic polarization scheme.

%The paper is structured as follows: First, we introduce the evolution for the dynamically polarized SERF comagnetometer and theoretically analyze the responses to spin perturbations in Sec.~\ref{Sec:Theory}. Next, we describe the experimental setup and operation procedures in Sec.~\ref{Sec:ExperSetup}. Subsequently, we present and discuss our experimental results in Sec.~\ref{Sec:ReADiss}. Finally, we conclude the paper with a summary of our findings in Sec.~\ref{Sec:Conclusion}.

\section{Result and Discussion}
\label{Sec:ReADiss}
\subsection{Basic Principle}
\begin{figure*}[htbp]
\centering   
\includegraphics[width=\linewidth]{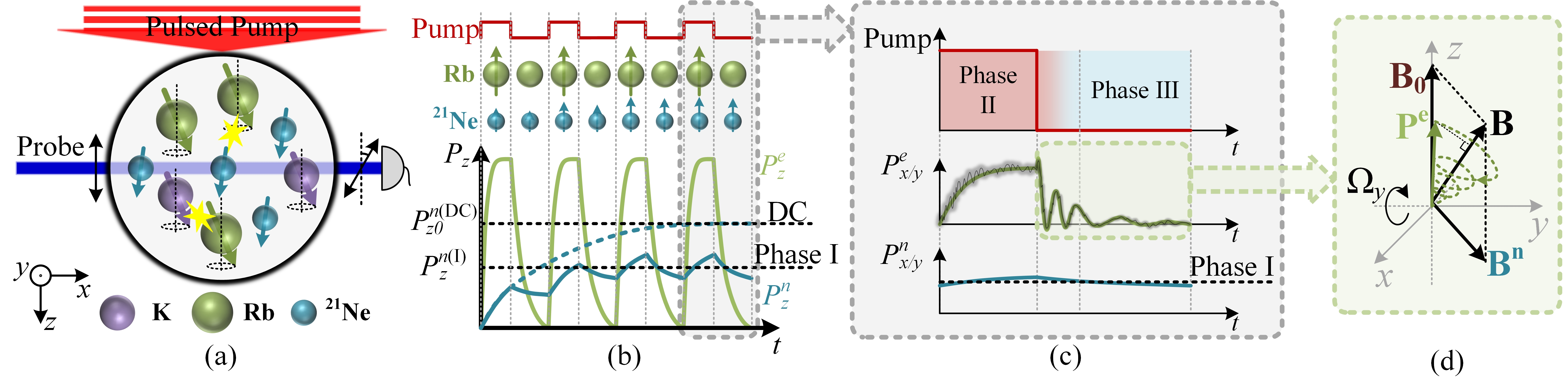}%
\caption{Basic operation of the dynamically polarized SERF comagnetometer based on pulsed pump light. (a) The pump and probe configurations. The pulsed pump light along $\hat{z}$ polarizes the K atoms directly. Rb atoms are polarized by K atoms and $^{21}$Ne atoms are hyperpolarized by Rb atoms through spin exchange. The probe light is utilized to measure the polarization of electron spins along $\hat{x}$ through optical rotation during the pump-on and pump-off phases.  (b) Under the same pumping rate, the process by which nuclear spins gradually polarize to a steady state under DC mode $P_{z0}^{n(\mathrm{DC})}$ and dynamic polarization $P_{z0}^{n(\mathrm{\uppercase\expandafter{\romannumeral1}})}$ (Phase \uppercase\expandafter{\romannumeral1}).  (c) The evolution of electron ($P_{x/y}^e$) and nuclear ($P_{x/y}^n$) spins transverse polarizations occurs as the pump light switches. The evolution of electron spins is designated as Phase \uppercase\expandafter{\romannumeral2} during the pump-on phase. After the light is turned off, when electron spins precess to a frequency-stable phase, the evolution is denoted as Phase \uppercase\expandafter{\romannumeral3}. Compared to the pump-on phase, the signal in the pump-off phase is not directly affected by the pump light, with polarization noise intended to be suppressed. The polarization of nuclear spins retains the steady-state value established in Phase \uppercase\expandafter{\romannumeral1}. It is important to note that the polarizations along sensor's $\hat{x}$ and $\hat{y}$ are not entirely identical, although they share the same form; therefore, we present them together.
(d) In the presence of external inputs, such as the inertial rotation along $\hat{y}$ ($\Omega_y$), the nuclear spin and the steady state of electron spin in Phase \uppercase\expandafter{\romannumeral2} develop non-zero components in the $x\mathrm{O}y$ platform. After the termination of the pump light, electron spins exhibit damped oscillation around the external magnetic field and the effective magnetic field of nuclear spins. }
\label{fig:Theory}
\end{figure*}

%We employ an acousto-optic modulator (AOM) to achieve the dynamic polarization via pulsed pump light,  In the pump-off phase, the polarization is intended to be suppressed owing to the signal being unaffected by the pump light directly. 
As shown in Fig.~\ref{fig:Theory}(a), a circularly polarized resonant light along $\hat{z}$ is used to polarize the K electron spins, which in turn polarize the optically thick Rb electron spins through spin-exchange, thereby improving polarization homogeneity. The probe light along $\hat{x}$ is used to measure the $\hat{x}$ component of electron spins polarization. The dynamics of coupled spin ensemble are reduced to a set of Bloch equations, which successfully describes the direct and spin-exchange optical pumping, spin relaxation, Fermi contact interaction, and spin perturbations. Since the rapid spin-exchange rate is significantly greater than the pulsed frequency, the evolution of electron spins in the cell can be simplified to a single effective equation, similar to that in DC mode~\cite{Wei2021Accurate}. The noble gas nuclear spins are polarized via hybrid spin-exchange optical pumping. Given the substantial difference in relaxation times between these two spin species, the alkali electron spins depolarize within tens of milliseconds after the pump light disappears, while the polarization of the noble gas nuclear spins remains approximately stable under dynamic polarization at several Hz. Based on the characteristics of alkali and noble gas spins, we divide the dynamic process into three phases to model the evolution of spin species.

The response of the alkali electron spins to spin perturbations depends on the perturbations themselves and the effective field produced by the noble gas nuclear spins through Fermi contact interaction. The polarization of the nuclear spins influences the entire process of the electron spins. Therefore, in Phase \uppercase\expandafter{\romannumeral1}, we aim to model the polarization of the nuclear spins, which remains nearly steady after polarization to a steady state, similar to that in DC mode. Upon switching the pump light, the longitudinal polarization of the electron spins rapidly polarizes and then depolarizes, while the nuclear spins gradually reach a steady state, as shown in Fig.~\ref{fig:Theory}(b). The polarization and depolarization dynamic of the noble gas nuclear spins are described by

\begin{align}
\begin{aligned}
% \left\{ \begin{array}{l}
P_z^n\left( t \right) \uparrow  &= P_{z0}^n\left( {1 - \exp \left\{ { - R_1^nt} \right\}} \right)\,,\\
P_z^n\left( t \right) \downarrow  &= P_{z0}^n\exp \left\{ { - R_1^nt} \right\}\,,
% \end{array} \right.
\end{aligned}
\end{align}
where $P_z^n$ is the longitudinal polarization of noble gas nuclear spins. $P_{z0}^n$ is the steady value of the polarization process and the initial value of the depolarization process. $R_1^n$ represents the longitudinal relaxation rate of nuclear spins. The contribution to $R_1^n$ induced by the pump light switching, which originates from the spin exchange rate with alkali metals ($R_{en}^{se}$), typically accounts for less than one-tenth of $R_1^n$. Therefore, we can reasonably assume that $R_1^n$ remains constant during the processes of polarization and depolarization. Dynamic polarization of nuclear spins reaches a steady state when the changes in the longitudinal polarization of nuclear spins induced by the switching process are balanced.
%In the polarization phase, the electron spin relaxation rate incorporates a non-zero pumping rate, making the polarization process more rapid than the depolarization.
Thus, the longitudinal polarization of noble gas nuclear spins is approximately proportional to the duty ratio ($D_r$) compared to that in the DC mode, when the pump rate is consistent with that in the pump phase, as given by ($P_{z}^{n(\mathrm{\uppercase\expandafter{\romannumeral1}})}=D_r P_{z0}^{n(\mathrm{DC})}$). Once the longitudinal polarization of the nuclear spins reaches a steady state, the $\hat{x}$ and $\hat{y}$ components of nuclear spins polarization remains essentially stable due to the long relaxation time and slow Larmor precession. 
%Once the longitudinal polarization of nuclear spin reaches equilibrium, its transverse polarization also remains steady.

As shown in Fig.\ref{fig:Theory} (c), in the pump-on phase, the transverse components, i.e., the $\hat{x}$ and $\hat{y}$ components of electron spin polarization reach steady states similar to the DC mode named Phase \uppercase\expandafter{\romannumeral2}. Once the pump is terminated, these components begin to exhibit damped oscillations along the combined magnetic field from the bias magnetic field and the effective magnetic field of nuclear spins, as shown in Fig.\ref{fig:Theory} (d). The entire polarization and depolarization of electron spins make the slowing down factor $Q$ time-dependent. Thus, the precession frequency changes in accordance with the variations in the longitudinal polarization of the electron spins, thereby introducing nonlinearity to the dynamic evolution. The numerical calculations of density matrix $\rho$ equation, which describes the evolution of Zeeman sublevels, make it difficult to obtain an analytical expression, and this hinders quantitative analysis. We find that when the polarization decays to a near-zero level, the precession frequency changes slowly. Therefore, the later phase of damped oscillation can be approximated as a linear evolution process, named Phase \uppercase\expandafter{\romannumeral3}.

\subsection{Experimental Setup}
\label{Sec:ExperSetup}
\begin{figure}[htbp]
\centering
\includegraphics[width=\linewidth]{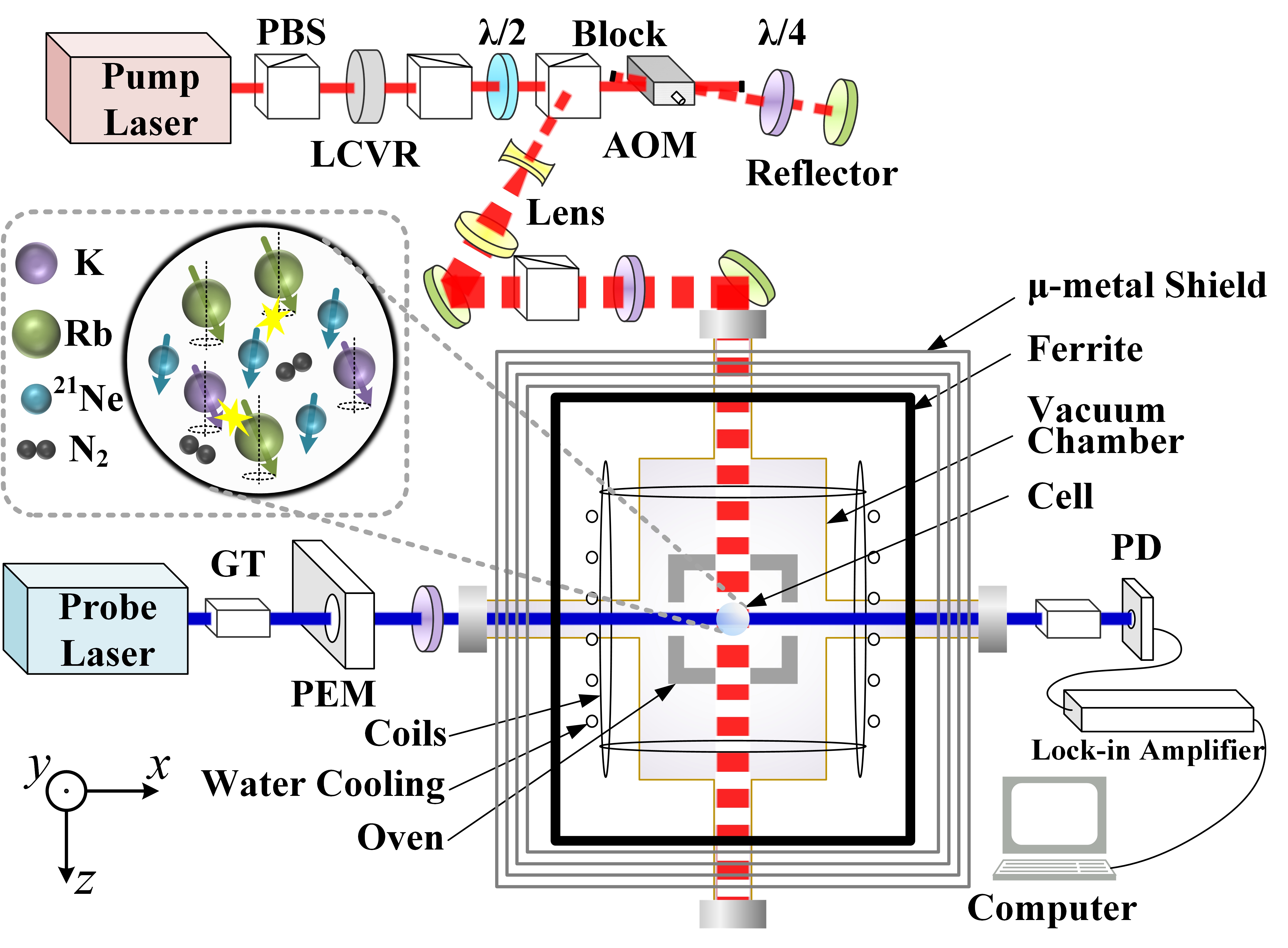}%
\caption{Experimental setup of K-Rb-$^{21}$Ne dynamically polarized SERF comagnetometer based on pulsed pump light. PBS, polarization beam splitter; LCVR, liquid crystal variable retarder; $\lambda$/2, half-wave plate; $\lambda$/4, quarter-wave plate; AOM, acousto-optic modulator; GT, Glan-Taylor prism; PEM, photoelastic modulator;
PD, photodetector.}
\label{fig:ppexpsetup}
\end{figure}

The experimental setup of our K-Rb-$^{21}$Ne dynamically polarized SERF comagnetometer utilizing pulsed pump light is illustrated in Fig.~\ref{fig:ppexpsetup}. A spherical cell made from GE180 glass with diameter of 14 mm is set at the center, which contains a small amount of potassium, a droplet of naturally abundant rubidium, with 3.24 amg $^{21}$Ne gas (70$\%$ isotope enriched) and 58 torr N$_2$ as quenching gas. The cell is heated to 453 K in a ceramic oven equipped with a set of twisted-pair wires, through which a 200 kHz high-frequency alternating current is passed. The polyether ether ketone vacuum chamber is used to reduce thermal convection. A ferrite and four-layer $\mu$-metal magnetic shield are used to shield the Earth's (or laboratory) magnetic field, providing a remanence less than 2 nT in three directions after degaussing,  which are further compensated by magnetic coils. Water cooling prevents high temperatures near the cell from affecting magnetic shielding performance. A circularly polarized resonant pump light generated by a fiber laser tuned to the D1 line of potassium (770.108 nm) polarizes the alkali-metal electrons along $\hat{z}$. A linearly polarized far-detuned probe light (795.501 nm)  generated by a distributed feedback diode laser propagating along $\hat{x}$ is used to measure the $\hat{x}$ component of electron spin polarization through optical rotation. The photoelastic modulator (PEM-100, Hinds Instruments) modulates the probe light to reduce low-frequency noise. 

We construct an acousto-optic modulator (AOM) (MT200-B100A0.5-800, AA Opto-Electronic) double-pass configuration to modulate the pump light at 11 Hz, aiming to achieve a higher isolation ratio. The control signal, provided by a waveform generator (33500B, Keysight Technologies), is utilized for switching the RF power amplifier (AMPB-B-34-10.500, AA Opto-Electronic) and triggering data acquisition. Since the temperature in our experiments is consistent, we only change the duty ratio of control signals from 40$\%$ to 60$\%$ in 5$\%$ increments. Thus, the longitudinal/transverse relaxation rates of electron spins in Phase $\mathrm{\uppercase\expandafter{\romannumeral2}}$ ($R_{1/2}^{e(\mathrm{\uppercase\expandafter{\romannumeral2}})}$) and Phase $\mathrm{\uppercase\expandafter{\romannumeral3}}$ ($R_{1/2}^{e(\mathrm{\uppercase\expandafter{\romannumeral3}})}$) at different duty ratios can be considered essentially the same. Therefore, for each duty ratio, we select 35 ms after the positive edge trigger as the steady state of Phase $\mathrm{\uppercase\expandafter{\romannumeral2}}$ and 20 ms - 35 ms after the negative edge trigger as the signal for Phase $\mathrm{\uppercase\expandafter{\romannumeral3}}$.

The entire apparatus is mounted on a single-axis rotary platform (not shown in Fig.~\ref{fig:ppexpsetup}). This enables the comagnetometer to rotate about the sensor's $\hat{z}$, thereby adjusting the projection of the Earth's rotation along sensor's $\hat{x}$ and $\hat{y}$. Consequently, we can utilize the Earth's rotation to calibrate the scale factor of the inertial rotation response.

\subsection{Magnetic Field Compensation}

The essence of self-compensation is that, when there is an external transverse magnetic field input, the effective magnetic field of nuclear spin generates an equal and opposite magnetic field to achieve compensation. This phenomenon enables the comagnetometer to maintain sensitivity to inertial rotation and anomalous forces while shielding the magnetic field interference. According to the theoretical analysis in the \textit{Magnetic Field Compensation} subsection of \textit{METHODS}, we investigate a magnetic field compensation method for dynamic polarization that leverages the steady state of Phase $\mathrm{\uppercase\expandafter{\romannumeral2}}$. By modulating square waves along sensor's $\hat{y}$ or $\hat{z}$ and comparing steady-state at high and low levels, we derive the first derivatives with respect to $B_y$ and $\delta B_z$. Applying another square wave along $\hat{z}$ with varying bias yields the second derivative with respect to $\delta B_z$. Setting these derivatives-the first with respect to $B_y$ and both the first and second with respect to $\delta B_z$-to zero achieves effective three-axis magnetic compensation, as in the DC mode~\cite{Li2023Magnetic,Gong2025Decoupling}. This compensation point constitutes a special point for nuclear spins throughout electron spin evolution, thereby enabling self-compensation across the pump-on phase (Phase $\mathrm{\uppercase\expandafter{\romannumeral2}}$) and pump-off phase (Phase $\mathrm{\uppercase\expandafter{\romannumeral3}}$).

\begin{figure}[htbp]
\centering
\includegraphics[width=\linewidth]{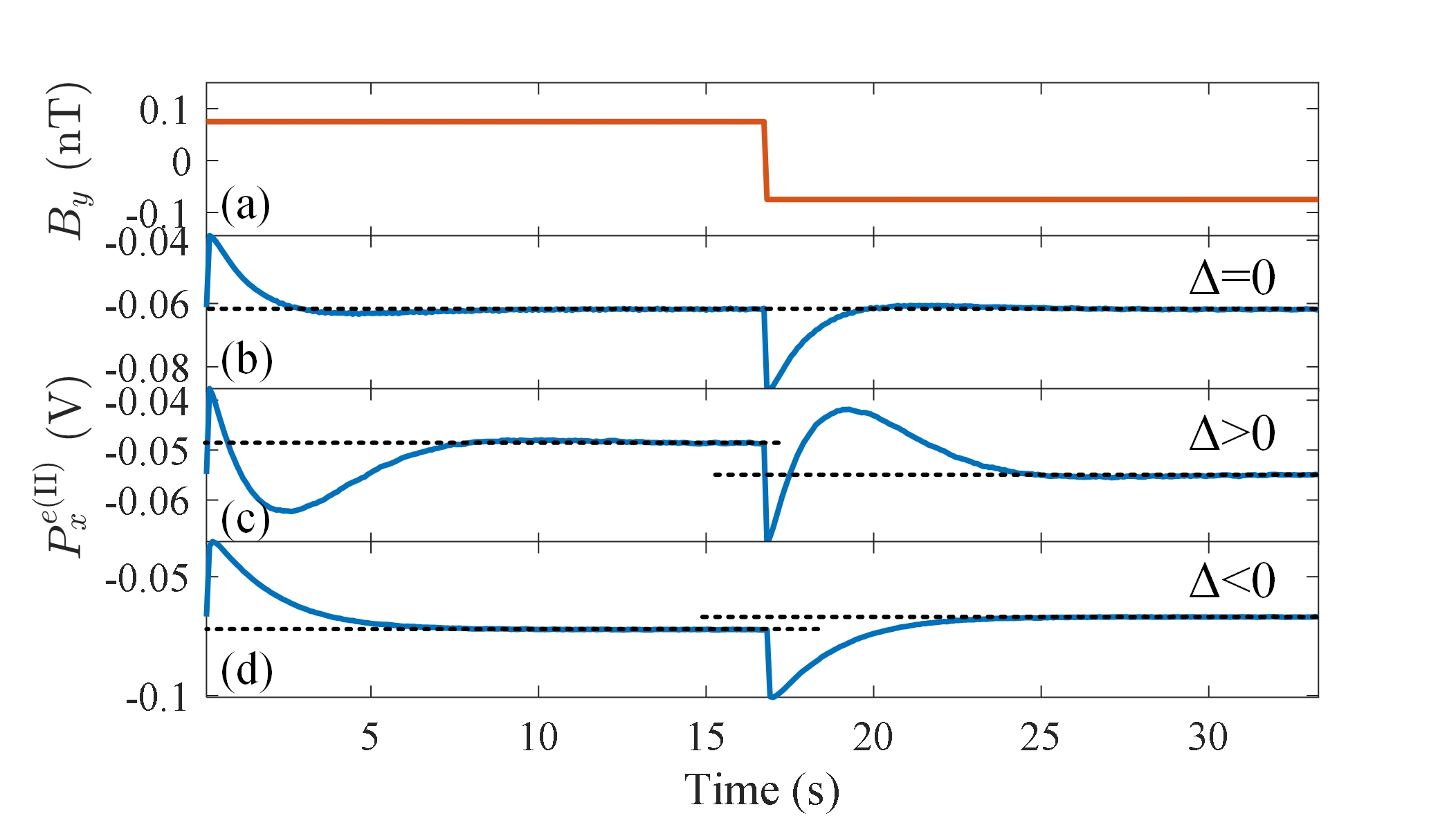}%
\caption{Procedure for zeroing $\delta B_z$. (a) A square wave modulation along $\hat{y}$ with a frequency of 30 mHz and an amplitude of 0.075 nT. (b)-(d) The steady state of Phase $\mathrm{\uppercase\expandafter{\romannumeral2}}$ at the longitudinal compensation point and at deviations in two directions from this point. By adjusting the magnetic field along $\hat{z}$, the steady states corresponding to high and low levels can be made consistent ($\Delta =0$), thereby achieving $\delta B_z=0$.}
\label{fig:magcompen}
\end{figure}

Fig.~\ref{fig:magcompen} illustrates the process of magnetic compensation along $\hat{z}$, which involves zeroing $\delta B_z$, defined as the difference between the magnetic field along $\hat{z}$ and the longitudinal compensation point. The applied square wave modulation along $\hat{y}$ with the frequency of 30 mHz and the amplitude of 0.075 nT is shown in Fig.~\ref{fig:magcompen}(a), while Fig.~\ref{fig:magcompen}(b)-(d) exhibit the steady state of Phase $\mathrm{\uppercase\expandafter{\romannumeral2}}$ at the compensation point and at deviations in two directions from this point, respectively. These states correspond to the differences in steady state values between high and low levels of the square wave $\Delta=0$, $\Delta>0$, and $\Delta<0$. According to our switching frequency of 11 Hz, one square wave period encompasses approximately 366 points. The dynamic process through which nuclear spin transitions to a steady state following the transverse magnetic field perturbation is obvious. By modulating $\delta B_z$ to achieve $\Delta=0$, the zeroing of $\delta B_Z$ is accomplished. For the zeroing of the magnetic fields along $\hat{x}$ and $\hat{y}$, a square wave is applied along $\hat{z}$, with the bias adjusted at $\delta B_z=0$  and slightly offset from it. Iterative regulation of the $\hat{x}$ and $\hat{y}$ magnetic fields ensures that the steady state differences between high and low levels are both zero at these two bias.

\subsection{Electron Damped Oscillation}

\begin{figure}[htbp]
\centering
\includegraphics[width=0.9\linewidth]{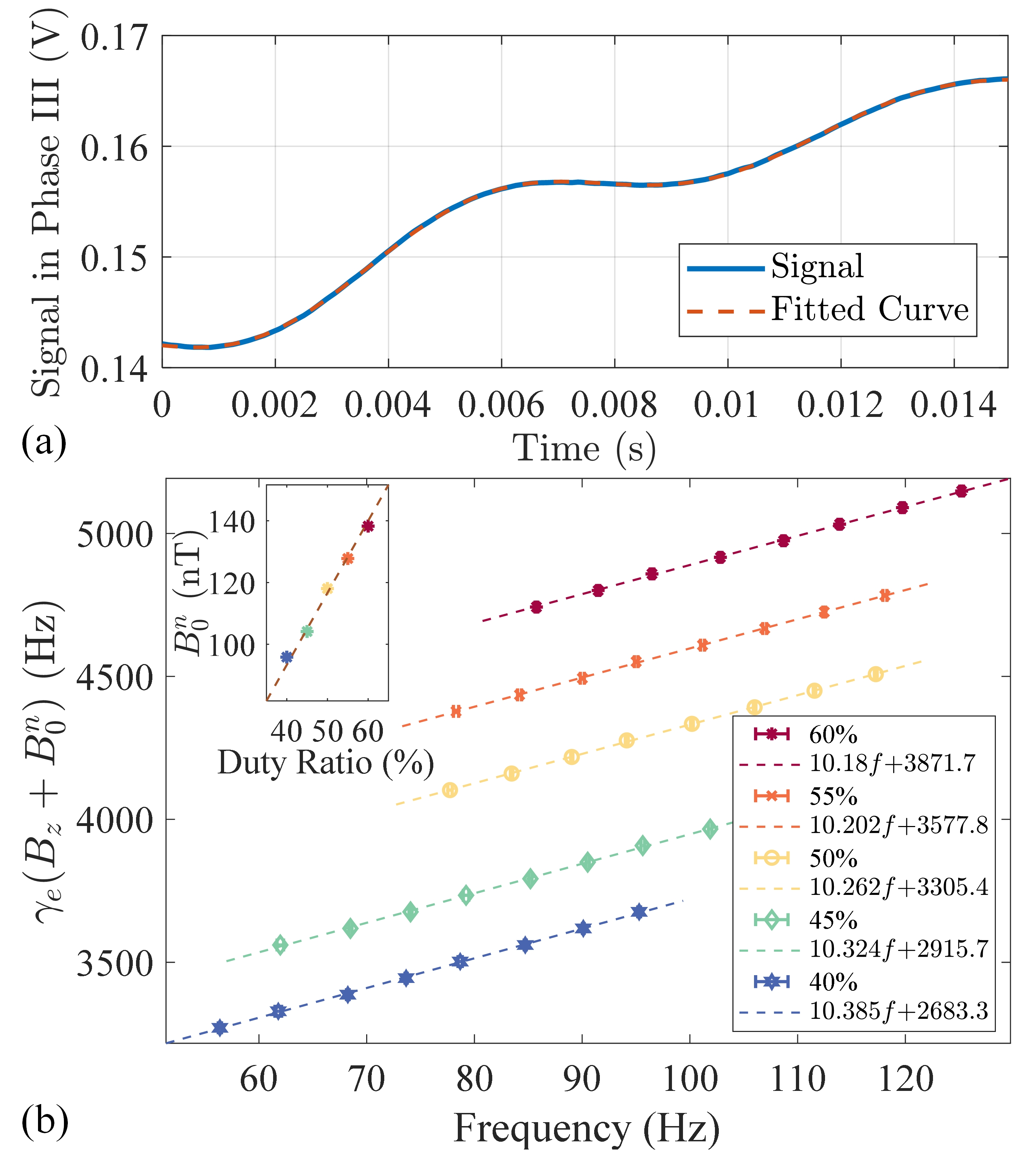}%
\caption{(a) One set of signals along with its fitting result. The fitting results using the fixed-frequency damped oscillation formula (Eq.~\ref{eq:freephi}) demonstrate an excellent agreement with the signal, with a root mean square error of $4.01 \times 10^{-5}$ V between the measured signal and the fitted curve.
(b)Relationship between the precession frequency and the longitudinal magnetic field. The slope is associated with the slowing down factor $Q^{(\mathrm{\uppercase\expandafter{\romannumeral3}})}$, which indicates that in the Phase $\mathrm{\uppercase\expandafter{\romannumeral3}}$ chosen for the experiments, the electron spin decays to less than 1/e of its initial value, while the precession frequency remains essentially stable. Inset: Relationship between the longitudinal nuclear effective magnetic field and duty ratio, along with the linear fitting using $y=kx$. This indicates that the longitudinal nuclear spin polarization is proportional to the duty ratio. }
\label{fig:BzfQ}
\end{figure}

The electron spin precession frequency in Phase $\mathrm{\uppercase\expandafter{\romannumeral3}}$ depends on the external longitudinal magnetic field, the nuclear spin effective magnetic field, and the slowing down factor, which is expressed by $f=\gamma_e (B_z+B_0^n)/Q^{(\mathrm{\uppercase\expandafter{\romannumeral3}})}$, where $f$ is the precession frequency. $\gamma_e$ represents the gyromagnetic ratio of alkali electron spins. $B_z$ is the longitudinal magnetic field. $B_0^n$ represents the longitudinal nuclear effective magnetic field.  $Q^{(\mathrm{\uppercase\expandafter{\romannumeral3}})}$ is the hybrid slowing down factor of K, $^{87}$ Rb and $^{85}$ Rb in Phase $\mathrm{\uppercase\expandafter{\romannumeral3}}$ which is related to the longitudinal polarization of electron spins. By varying $B_z$ near the compensation point, the relationship between frequency and magnetic field can be obtained, thereby determining the slowing down factor for Phase $\mathrm{\uppercase\expandafter{\romannumeral3}}$ and the nuclear spin effective magnetic field at different duty ratios. Since our focus here is on precession frequency, there is no need to specify a particular fitting phase. Thus, the fitting formula is
\begin{align}
\begin{aligned}
P_x^{e(\mathrm{\uppercase\expandafter{\romannumeral3}})}=&\exp\left\{ -\frac{R_2^{e(\mathrm{\uppercase\expandafter{\romannumeral3}})}}{Q^{(\mathrm{\uppercase\expandafter{\romannumeral3}})}} t \right\}A\sin (2\pi f t +\phi)\\
&+B\exp\left\{ -\frac{R_1^{e(\mathrm{\uppercase\expandafter{\romannumeral3}})}}{Q^{(\mathrm{\uppercase\expandafter{\romannumeral3}})}} t \right\}+C\,,\label{eq:freephi}
\end{aligned}
\end{align}
where ${R_{2/1}^{e(\mathrm{\uppercase\expandafter{\romannumeral3}})}}/{Q^{(\mathrm{\uppercase\expandafter{\romannumeral3}})}}$, $A$, $f$, $\phi$, $B$, $C$ are all fitting parameters.  One representative set of signals and its fitting result is shown in Fig.~\ref{fig:BzfQ}(a). The relationships between magnetic field and precession frequency under various duty ratios are summarized in Fig.~\ref{fig:BzfQ}(b), where the slope represents the hybrid slowing down factor $Q^{(\mathrm{\uppercase\expandafter{\romannumeral3}})}$. The slowing down factors for the five sets are essentially consistent, with an average value of 10.27, corresponding to a longitudinal electron polarization of approximately 18 $\%$~\cite{Xu2021Fast}. Under the same pumping rate, the longitudinal polarization rate in DC mode is about 77 $\%$. Thus, in the duration we used for Phase $\mathrm{\uppercase\expandafter{\romannumeral3}}$, the longitudinal electron polarization has decayed to below 1/e of the initial polarization. Therefore, the polarization decays relatively slowly in this duration. From the fitting relationship between $B_0^n$ and duty ratio in the inset of Fig.~\ref{fig:BzfQ}, it is evident that the nuclear spin effective polarization exhibits a basically linear relationship with the duty ratio, consistent with our analysis.

\subsection{Response to Inertial Rotation}

\begin{figure*}[htbp]
\centering
\includegraphics[width=\linewidth]{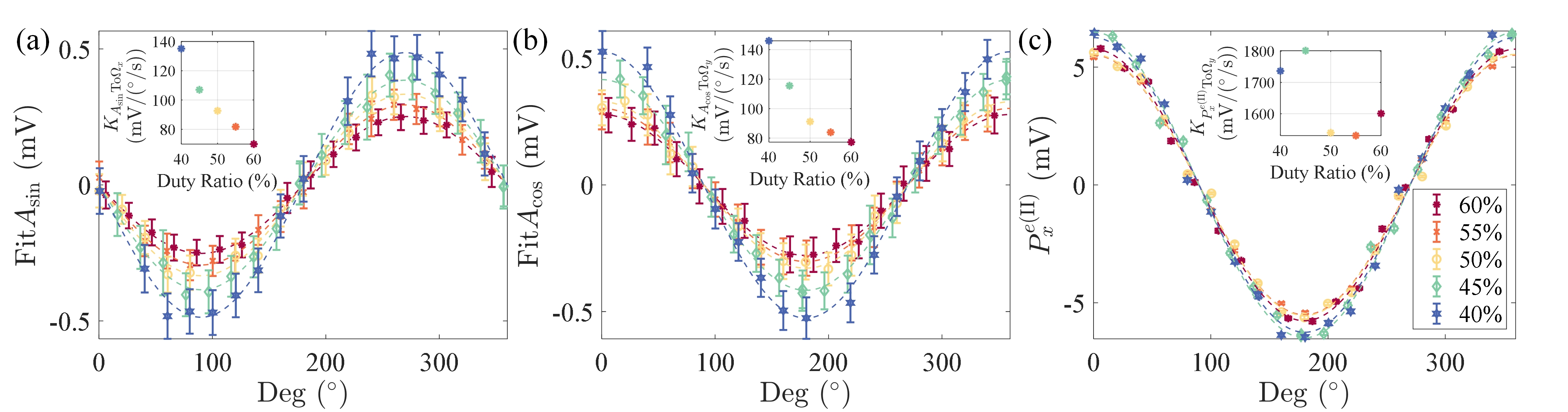}%
\caption{Response to Earth's rotation. At the initial position of the rotary platform (where Deg = 0$^{\circ}$), the sensor's $\hat{y}$ is aligned parallel to the north-south direction. 
As the rotary platform completes a full rotation about the sensor's $\hat{z}$ (from 0$^{\circ}$ to 360$^{\circ}$), it can be observed that the fitting parameter $A_{\sin}$ (a) for the sinusoidal term of Eq.~\ref{eq:fitAsincos} reflect the variation trend of the projection of Earth's rotation along the sensor's $\hat{x}$ ($\Omega_x$), whereas the fitting parameter $A_{\cos}$ (b)  for the cosine term and the steady state $P_x^{e(\mathrm{\uppercase\expandafter{\romannumeral2}})}$ of Phase $\mathrm{\uppercase\expandafter{\romannumeral2}}$ (c) reflect the variation pattern of the projection of Earth's rotation along the sensor's $\hat{y}$ ($\Omega_y$).
Inset: the relationship between the scale factor of inertial rotation and duty ratio. The scale factors for inertial rotation of $A_{\sin}$ and $A_{\cos}$ decrease with increasing duty ratio, while the scale factor for $P_x^{e(\mathrm{\uppercase\expandafter{\romannumeral2}})}$ shows little correlation with the duty ratio.}
\label{fig:Omegatest}
\end{figure*}

Earth's rotation is used to calibrate the response to inertial rotation. With the sensor's $\hat{y}$ oriented in the north-south direction as the initial position, the comagnetometer is rotated around the sensor's $\hat{z}$, with measurements conducted every 20$^\circ$. To eliminate fluctuations in bias, we employ a differential measurement strategy using two test points spaced 180$^\circ$ apart as one group, resulting in the results presented having a bias of zero. Here, we utilize the formula that account for the special fitting phase as

\begin{align}
\begin{aligned}
P_x^{e(\mathrm{\uppercase\expandafter{\romannumeral3}})}=&\exp\left\{ -\frac{R_2^{e(\mathrm{\uppercase\expandafter{\romannumeral3}})}}{Q^{(\mathrm{\uppercase\expandafter{\romannumeral3}})}} t \right\} \\
&\times\left[A_{\sin}\sin (2\pi f t +\phi_\Omega)+A_{\cos}\cos (2\pi f t +\phi_\Omega) \right]\\
&+B\exp\left\{ -\frac{R_1^{e(\mathrm{\uppercase\expandafter{\romannumeral3}})}}{Q^{(\mathrm{\uppercase\expandafter{\romannumeral3}})}} t \right\}+C\,,\label{eq:fitAsincos}
\end{aligned}
\end{align}
where $\phi_\Omega$ is the special fitting phase for each experimental condition. $A_{\sin}$ and $A_{\cos}$ are expected to be associated with inertial rotation along sensor's $\hat{x}$ ($\Omega_x$) and $\hat{y}$ ($\Omega_y$), respectively. The fitting results are illustrated in Fig.~\ref{fig:Omegatest}(a)-(b). With the sensor's $\hat{y}$ oriented in the north-south direction as the initial angle, during a full rotation of the rotary platform around $\hat{z}$, $\Omega_x$ exhibits a sinusoidal form, while $\Omega_y$ exhibits a cosine form. It is evident that our fitting result $A_{\sin}$ also takes a sinusoidal form, corresponding to $\Omega_x$, and $A_{\cos}$ takes a cosine form, corresponding to $\Omega_y$. Furthermore, we examined the steady state of Phase $\mathrm{\uppercase\expandafter{\romannumeral2}}$ ($P_x^{e(\mathrm{\uppercase\expandafter{\romannumeral2}})}$) as shown in Fig.~\ref{fig:Omegatest}(c), which exhibits a cosine form corresponding to $\Omega_y$, and this result aligns well with the response observed under DC mode~\cite{wei2022constraints}. Each point represents the average of 190 measurements, with error bars indicating the standard deviation. The larger standard deviation in Phase $\mathrm{\uppercase\expandafter{\romannumeral3}}$ is attributed to magnetic field disturbances near the cell, (which is discussed in the \textit{Sensitivity} subsection). The insets in Fig.~\ref{fig:Omegatest} illustrate the relationship between the scale factor of inertial rotation and the duty ratio. It is evident that in Phase $\mathrm{\uppercase\expandafter{\romannumeral3}}$, the scale factor decreases as the duty ratio increases, which is consistent with our simulation result in Fig.~\ref{fig:FigKomega}. In contrast, the scale factor in Phase $\mathrm{\uppercase\expandafter{\romannumeral2}}$ exhibits little dependence on the duty ratio as the steady state longitudinal electron spin polarization and relaxation, which determine the scale factor in the DC mode, remain almost unaffected at different duty ratios.

\subsection{Suppression of Low-Frequency Magnetic Field}
The suppression of transverse DC magnetic fields and low-frequency magnetic disturbances is one of the primary advantages of the self-compensation comagnetometer, which is essential for  low-frequency inertial rotations or pseudomagnetic fields detection. Subsequently, we verify the low-frequency magnetic field suppression capability under dynamic polarization. A low-frequency magnetic field along sensor's $\hat{x}$ or $\hat{y}$ (ranging from 0.005 Hz to 1 Hz) is applied through the magnetic compensation coils. We utilize Eq.~\ref{eq:fitAsincos} for fitting. Thus, the results of the $A_{\cos}$ can be directly compared with the steady state of the Phase $\mathrm{\uppercase\expandafter{\romannumeral2}}$, corresponding to $P_x^e$ in DC mode, whereas the $A_{\sin}$ corresponds to $P_y^e$ of that. 

\begin{figure*}[htbp]
\centering
    \includegraphics[width=0.8\linewidth]{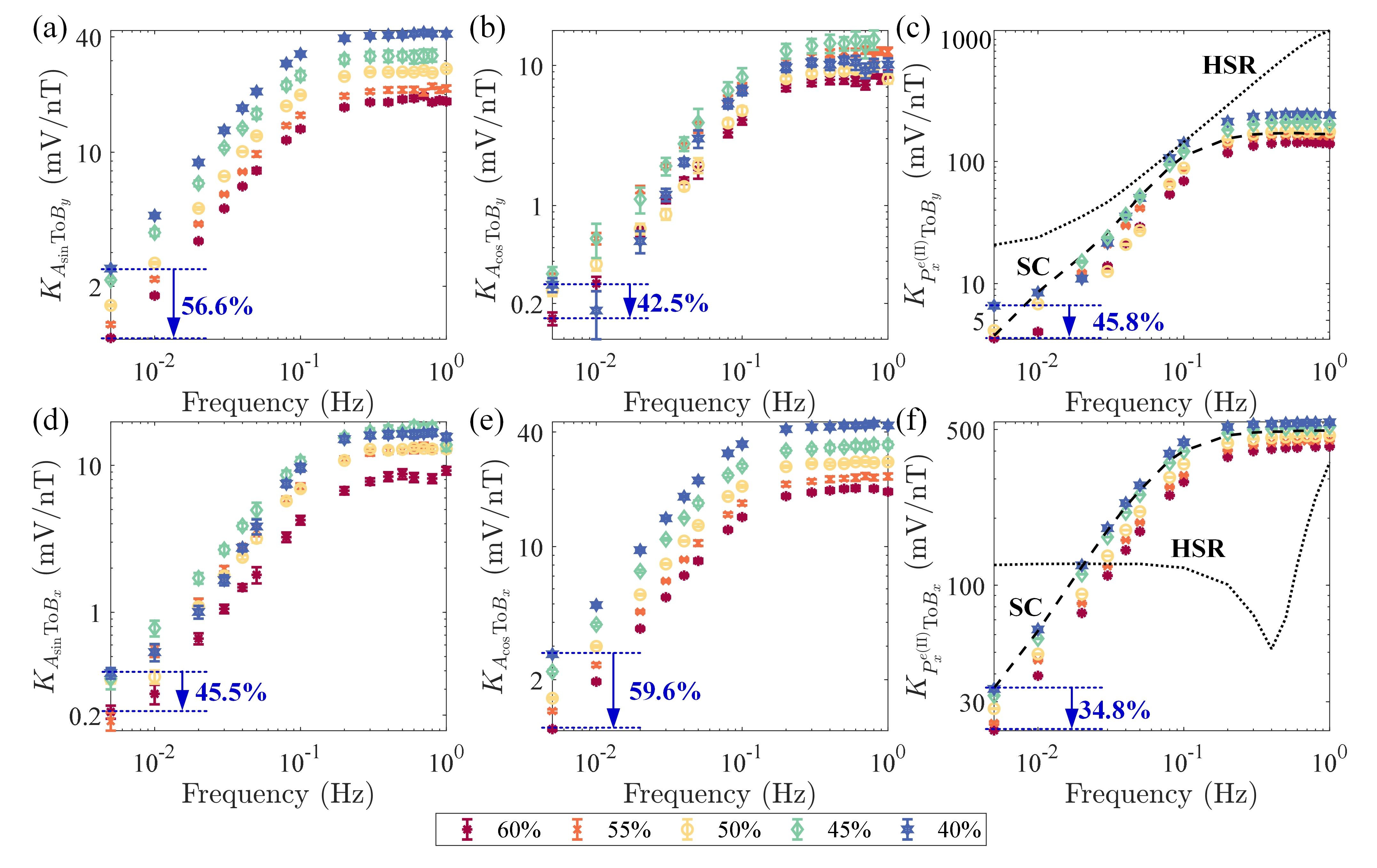}%
\caption{Response to low-frequency magnetic field. (a)-(c) represent the scale factor for $A_{\sin}$, $A_{\cos}$, and $P_x^{e(\mathrm{\uppercase\expandafter{\romannumeral2}})}$  in response to the low-frequency magnetic field along sensor's $\hat{y}$, respectively, while (d)-(f) correspond to those for the magnetic field along sensor's $\hat{x}$. Larger scale factor signifies greater response, corresponding to reduced suppression capability.
All the frequency responses within the tested frequency range exhibit high-pass filter characteristics, with the suppression of magnetic fields diminishing above 0.1 Hz. 
As the duty ratio increases, the fitting parameters $A_{\sin}$ and $A_{\cos}$, along with the steady state $P_x^{e(\mathrm{\uppercase\expandafter{\romannumeral2}})}$  of Phase $\mathrm{\uppercase\expandafter{\romannumeral2}}$, exhibit an overall decreasing response to the low-frequency magnetic fields along $\hat{x}$ and $\hat{y}$. By increasing the duty ratio from 40 $\%$ to 60 $\%$, the low-frequency magnetic field responses of $A_{\sin}$ and $A_{\cos}$ are suppressed by an average of 51.1 $\%$. The two black lines in  (c) and (f) illustrate a comparison of the response in the self-compensation (SC) mode and the hybrid spin resonance (HSR) mode with $B_z = -B_0^n$. It is evident that, in the self-compensation mode, the low-frequency magnetic field is significantly suppressed.
In (b), (c), and (d), the outliers that deviate from the pattern are attributed to system drift during long-time experiments, which causes a shift in the compensation point.
%Moreover, the deviations in $A_{\sin}$ and $A_{\cos}$ are more pronounced because changes in the system state affect not only the compensation point but also the decoupling phase.
}
\label{fig:freresptest}
\end{figure*}

As illustrated in the first row of Fig.~\ref{fig:freresptest} (Fig.~\ref{fig:freresptest}(a)-(c), the scale factors for the responses of the $A_{\sin}$, $A_{\cos}$, and $P_x^{e(\mathrm{\uppercase\expandafter{\romannumeral2}})}$ to different frequencies of $B_y$ magnetic fields are presented. The second row depicts the responses to $B_x$. It is evident that the $A_{\sin}$ term exhibits a larger response to $B_y$ than to $B_x$, whereas the $A_{\cos}$ term and $P_x^{e(\mathrm{\uppercase\expandafter{\romannumeral2}})}$ demonstrate higher responses to $B_x$. Moreover, a decreasing trend in the response coefficients is observed with increasing duty ratio, indicating an enhanced capability to suppress magnetic fields. However, irregularities are noted in Fig.~\ref{fig:freresptest}(b) and (d), likely attributable to system drift causing slight deviations from the compensation point. Similarly, Fig.~\ref{fig:freresptest}(c) shows several anomalous data points at low frequencies, also influenced by system drift. As drift leads to corresponding shift in the decoupling phase $\phi_\Omega$ and the inherently smaller responses of the $A_{\sin}$ to $B_x$ and the $A_{\cos}$ term to $B_y$, the fitting results for Phase $\mathrm{\uppercase\expandafter{\romannumeral3}}$ present greater deviation. Overall, raising the duty ratio from 40 \% to 60 \% results in an average suppression of 51.1 $\%$ of low-frequency magnetic fields during the pump-off phase. 

In Fig.~\ref{fig:freresptest}(c) and (f), the black dashed lines represent the self-compensation mode, while the dotted lines denote the hybrid spin resonance mode with $B_z$ adjusted to $B_z = -B_0^n$. It is evident that the self-compensation mode demonstrates superior efficacy in suppressing the low-frequency magnetic field. The measurements in Phase $\mathrm{\uppercase\expandafter{\romannumeral3}}$ are based on electron spin precession, where alterations in the magnetic field induce changes in the precession frequency. As a result, a direct comparison of $A_{\sin}$ and $A_{\cos}$ between these two modes is not feasible. However, it is apparent that the lineshape of $A_{\sin}$ and $A_{\cos}$ is consistent with that observed in the self-compensation mode.

%Overall, the results reveal a consistent trend wherein increasing the duty ratio strengthens the magnetic field suppression capability.

% \begin{widetext}
% \begin{equation}\label{eqn-fit-fun}
% P_x^e = {{\rm{e}}^{ - t/{T_2}}}\left[ {A\sin \left( {2\pi ft + \varphi } \right) + D\cos \left( {2\pi ft + \varphi } \right)} \right] + B{{\rm{e}}^{ - t/{T_1}}} + C\,,
% \end{equation}
% \end{widetext}

\subsection{Suppression of Pump Light Fluctuation}

\begin{figure}[htbp]
\centering
\includegraphics[width=0.9\linewidth]{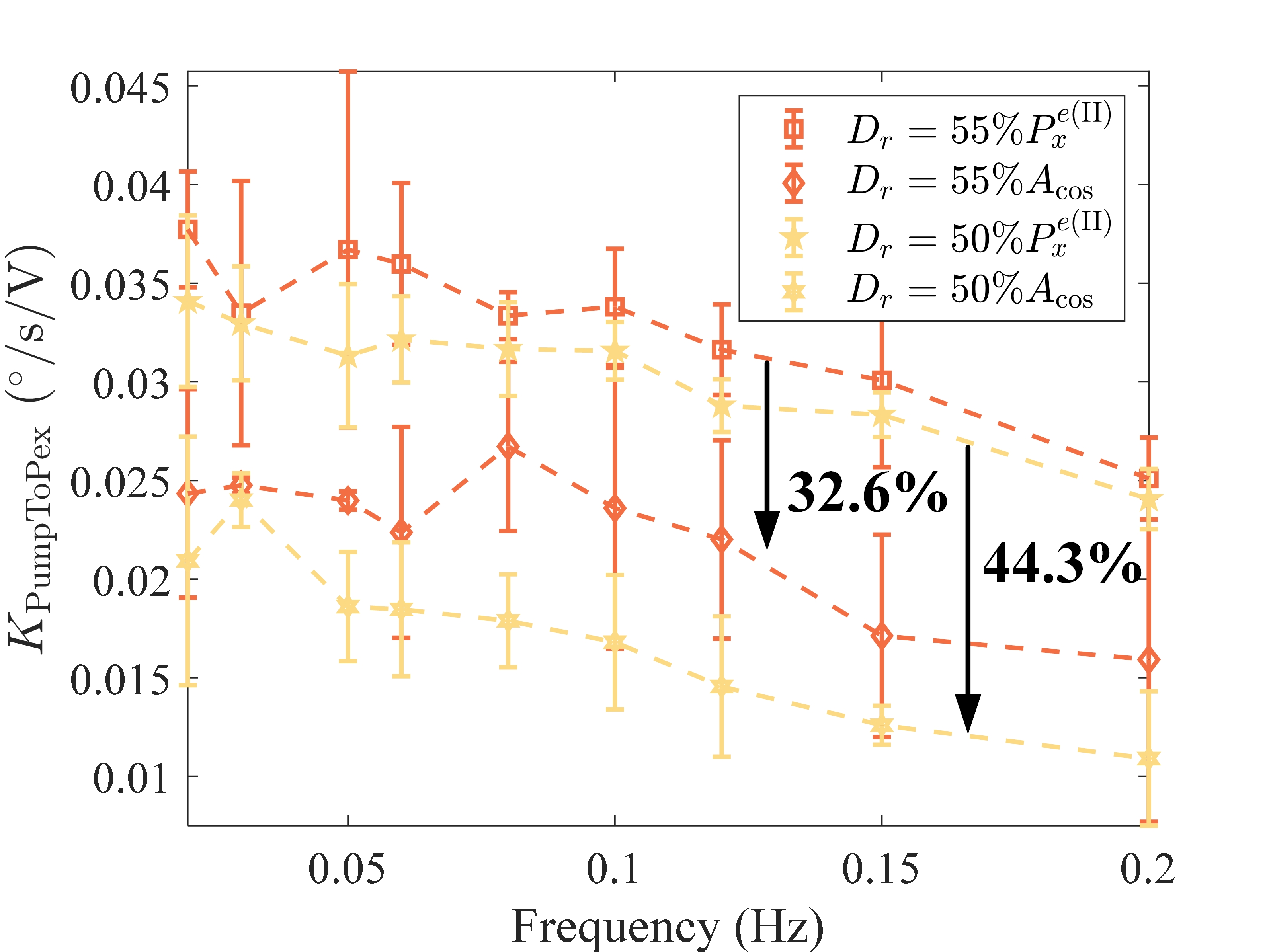}%
\caption{The relative response coefficients of $A_{\cos}$ and $P_x^{e(\mathrm{\uppercase\expandafter{\romannumeral2}})}$ to fluctuations in light intensity under the duty ratio of 50$\%$ and 55$\%$. Compared to Phase $\mathrm{\uppercase\expandafter{\romannumeral2}}$ with the pump light on, Phase $\mathrm{\uppercase\expandafter{\romannumeral3}}$ with the pump light off exhibits a clear suppression of the pump light fluctuations by an average of 38.5 $\%$. }
\label{fig:pumpFlu}
\end{figure}

The dynamic polarization scheme based on pulsed pump light is originally proposed to reduce polarization noise. Fluctuations in the pumping light intensity not only directly induce fluctuations in electron polarization but also cause signal fluctuations due to light shift and misalignment. Therefore, we modulate the pumping light intensity by varying the driving voltage of the liquid crystal variable retarder to evaluate the response of pump light intensity fluctuation. The corresponding light intensity fluctuations are detected by the photodiode and converted to voltage signals via the photodiode amplifier. According to Ref.~\cite{XU2024Quantification}, for each test frequency, we vary five modulation amplitudes, calculate the slope of the light intensity fluctuations (with the unit of V) against $A_{\cos}$ and $P_x^{e(\mathrm{\uppercase\expandafter{\romannumeral2}})}$ (with the unit of mV), divide this by the scale coefficient of $A_{\cos}$ and $P_x^{e(\mathrm{\uppercase\expandafter{\romannumeral2}})}$ with respect to $\Omega_y$ (in mV/($^\circ$/s)), and denote it as the relative response coefficient to pump light intensity fluctuation. Fig.~\ref{fig:pumpFlu} presents the results for 55 $\%$ and 50 $\%$ duty ratios. Due to the limitation in the pump light switching frequency and the observation that the response to light intensity fluctuations in the low-frequency range decreases as the frequency increases~\cite{XU2024Quantification}, we measure the response within 0.02 Hz - 0.2 Hz. It is evident that, under both duty ratios, the response coefficient of $A_{\cos}$ is smaller, indicating a suppression to pump light fluctuations during the pump-off phase (Phase $\mathrm{\uppercase\expandafter{\romannumeral3}}$). The overall average of all collected data yielded a value of 38.5 $\%$.

\subsection{Sensitivity}
\label{subsec:sensitivity}
\begin{figure}[htbp]
\centering
\includegraphics[width=1\linewidth]{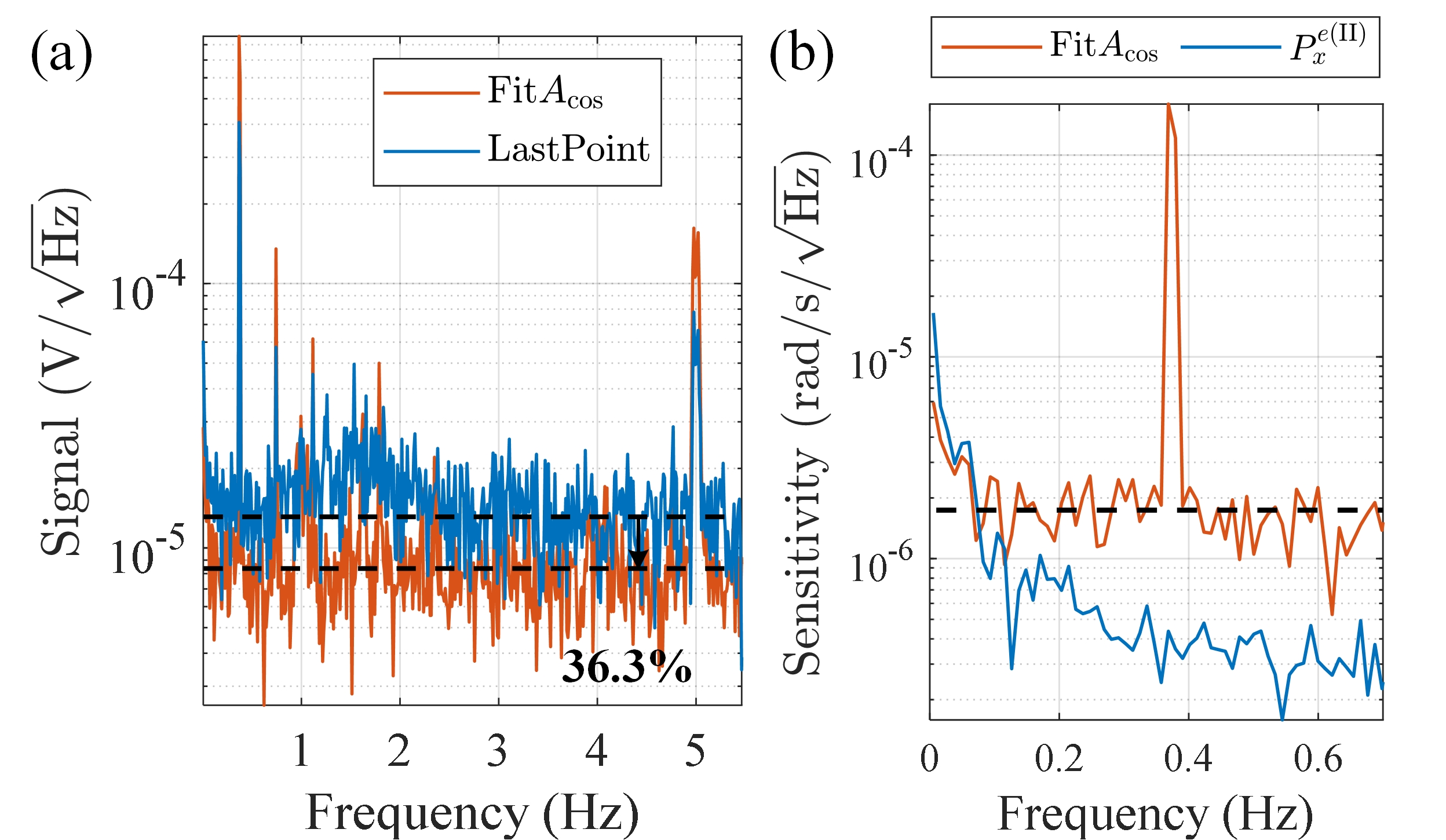}%
\caption{Signal and sensitivity comparison. (a) The noise floor comparison of $A_{\cos}$ and the last point in Phase $\mathrm{\uppercase\expandafter{\romannumeral3}}$. The dashed lines represent the mean value averaged over 3 Hz - 4 Hz. Compared to directly utilizing the last point of Phase $\mathrm{\uppercase\expandafter{\romannumeral3}}$, the  noise floor derived from fitting parameter $A_{\cos}$ is reduced by 36.3 $\%$. (b) The sensitivity comparison of $A_{\cos}$ and $P_x^{e(\mathrm{\uppercase\expandafter{\romannumeral2}})}$. Below 0.1 Hz, the sensitivity derived from the fitting parameter $A_{\cos}$ in Phase $\mathrm{\uppercase\expandafter{\romannumeral3}}$ is better than the steady state of Phase $\mathrm{\uppercase\expandafter{\romannumeral2}}$ (which represents DC mode). Above 0.1 Hz, however, the sensitivity of $A_{\cos}$ is constrained by the background noise.  }
\label{fig:sensitivity}
\end{figure}

The signals we obtain, denoted as $S$, typically comprise the quantities of interest $\Omega_y$ , noise components that are amplifiable $N_a$  (e.g., magnetic noise, pump light fluctuation), and detection backgrounds $N_{\mathrm{na}}$  that remain unaffected by increases in scale factor(such as photon shot noise and spin projection noise). The signal we measured can be expressed by
\begin{align}
    S=K_\Omega(\Omega_y+N_a)+N_{\mathrm{na}}\,,
\end{align}
where $K_\Omega$ represents the scale factor of $\Omega_y$. Once we calibrate the $\Omega_y$ through Earth's rotation calibration or other method, the sensitivity is defined by 
\begin{align}
    \frac{\Delta S}{K_\Omega}=\sqrt{(\Delta\Omega_y)^2 +(\Delta N_a)^2+\frac{(\Delta N_{\mathrm{na}})^2}{K_\Omega^2}}\,,
\end{align}
where $\Delta X$ represents the standard deviation of $X$. As $K_\Omega$ decreases, the effects of background noise become more pronounced. Under dynamic polarization scheme, the signal is measured after the pump light is turned off and the longitudinal electron spin polarization decays to nearly zero, which results in a smaller scale factor compared to DC mode (as in the insets of Fig.~\ref{fig:Omegatest}).  In the dynamic polarization scheme, the fitting method employed reduces noise floor (black dashed lines) by approximately 36.3 $\%$ compared with the signal attained by the last point of Phase $\mathrm{\uppercase\expandafter{\romannumeral3}}$ as shown in Fig.~\ref{fig:sensitivity}(a), and there is no amplification effect from the pumping light on probe noise (as reported in the Ref.~\cite{XU2024Quantification}). However, due to the small scale factor, the noise becomes more evident. Compared with the steady state of Phase $\mathrm{\uppercase\expandafter{\romannumeral2}}$ which represents the DC mode, the sensitivity of the 454.55 s data with 0.1 Hz binsize (Fig.~\ref{fig:sensitivity}(b)) shows that at very low frequencies ($<$0.1 Hz), dynamic polarization scheme improves sensitivity by suppressing pumping light fluctuation response. Nonetheless, above 0.1 Hz, the sensitivity tends to converge with the background noise as the black dashed line calculated by the averaged noise floor between 3 Hz to 4 Hz.

Due to the damped oscillation mechanism of dynamic polarization scheme, which heavily relies on magnetic field stability, it introduces additional noise peaks compared to continuous scheme, e.g., the noise peak around 0.37 Hz in Fig.~\ref{fig:sensitivity}(b). These noises originate from magnetic field fluctuations near the cell possible caused by electric heating. Constrained by the damped oscillation frequency at the compensation point, current pump switching frequency is 11 Hz, which results in some high-frequency noises in the system being shifted to lower frequencies, thereby impacting overall system performance. Therefore, future improvements should focus on optimizing background noise, enhancing magnetic field stability near the cell, and replacing fitting methods with other predictive approaches to improve the performance of dynamic polarization sensors.

\section{Conclusion}
\label{Sec:Conclusion}
Motivated by the potential of dynamic manipulation to suppress technical noise and overcome quantum noise limit, we investigate a K-Rb-$^{21}$Ne dynamic polarization scheme based on pulsed pump light. For quantitative analysis, we develop three phases to fully characterize the evolution of the alkali-noble-gas hybrid spin ensemble under dynamic polarization. A method for magnetic field compensation, based on steady-state responses in the pump-on phase, is explored. We focus on the response to inertial rotation and low-frequency magnetic fields at the compensation point. Experiments are conducted to evaluate the response of the coupled spin ensemble to various longitudinal magnetic fields, Earth's rotation, low-frequency magnetic fields, and pump light fluctuations at different duty ratios. Experimental verification demonstrates the effectiveness of the proposed method in characterizing damped oscillation dynamic phenomena, which will be helpful in analyzing the dynamic polarization evolution in other hybrid spin-based sensors.

Overall, the dynamic polarization scheme shows significant potential for suppressing polarization noise (of about 38.5 $\%$ in our experiments). By modulating the duty ratio, we achieve a 51.1 \% suppression of the low-frequency magnetic field response, thereby demonstrating the self-compensation capability under dynamic polarization. However, the damped oscillation mechanism imposes higher demands on magnetic field stability, probe noise, and other factors. These could be further improved in the future by integrating multi-reflection cavities~\cite{Sheng2013Subfemtotesla,ZHENG2025Enhancing}, self-differential measurements~\cite{zhou2025SERF}, and squeezing or entanglement~\cite{heng2025enhanced,kong2020measurement} to reduce probe noise. Laser heating serves as a viable substitute for electric heating to minimize magnetic field effects around the cell~\cite{zhao2022ultra,Li2023Laser}. Furthermore, physics-informed neural networks~\cite{chen2021physics, norambuena2024physics} and estimation algorithms~\cite{Smith2024Adaptive,nolan2021machine,pei2024real} could help discover novel schemes for signal extraction. In combination with optimization algorithms~\cite{xiao2023quantum,sharma2022comprehensive}, the dynamic polarization scheme is expected to enhance long-term stability and low-frequency performance, offering promising applications in high-precision inertial navigation and new physics explorations, such as the Fifth force and axion-like particles.

\begin{widetext}
\section*{Methods}
\label{Sec:Theory}

The effective magnetic field produced by nuclear spins is crucial for the whole process dynamics of electron spins. Thus, in Phase \uppercase\expandafter{\romannumeral1}, we establish an equivalent electron spins polarization state to assist in analyzing nuclear spins polarization, which ensures the nuclear spin longitudinal polarization corresponding to the product of that of DC mode under the same pump rate and the duty ratio of the pulsed sequence. The dynamic equations in Phase \uppercase\expandafter{\romannumeral1} is

\begin{align}
    \begin{aligned}\label{eqn-Phase1}
\frac{\partial {{\bf{P}}^{\bf{e}(\mathrm{\uppercase\expandafter{\romannumeral1}})}} }{\partial t} =& \frac{{{\gamma _e}}}{{{Q^{(\mathrm{\uppercase\expandafter{\romannumeral1}})}}}}({\bf{B}} + {{\bf{L}}^{(\mathrm{\uppercase\expandafter{\romannumeral1}})}} + \lambda M_0^n{{\bf{P}}^{\bf{n}}}) \times {{\bf{P}}^{\bf{e}(\mathrm{\uppercase\expandafter{\romannumeral1}})}} - {\bf{\Omega }} \times {{\bf{P}}^{\bf{e}(\mathrm{\uppercase\expandafter{\romannumeral1}})}} + \frac{{{R_p^{(\mathrm{\uppercase\expandafter{\romannumeral1}})}} {{\bf{S}}_{\bf{p}}} + R_{se}^{ne}{{\bf{P}}^{\bf{n}}}}}{Q^{(\mathrm{\uppercase\expandafter{\romannumeral1}})}} - \frac{{\{ R{{_2^e}^{(\mathrm{\uppercase\expandafter{\romannumeral1}})} },R{{_2^e}^{(\mathrm{\uppercase\expandafter{\romannumeral1}})} },R{{_1^e}^{(\mathrm{\uppercase\expandafter{\romannumeral1}})}}\} }}{Q^{(\mathrm{\uppercase\expandafter{\romannumeral1}})}}{{\bf{P}}^{\bf{e}}}^{(\mathrm{\uppercase\expandafter{\romannumeral1}})} \,,\\
\frac{{\partial {{\bf{P}}^{\bf{n}}}}}{{\partial t}} =& {\gamma _n}({\bf{B}} + \lambda M_0^e{{\bf{P}}^{\bf{e}}}) \times {{\bf{P}}^{\bf{n}}} - {\bf{\Omega }} \times {{\bf{P}}^{\bf{n}}}  + R_{se}^{en}{{\bf{P}}^{\bf{e}}}^{(\mathrm{\uppercase\expandafter{\romannumeral1}})}  - \{ R_2^n,R_2^n,R_1^n\} {{\bf{P}}^{\bf{n}}}\,,
    \end{aligned}
\end{align}
where ${\bf{P}}^{\bf{e}(\mathrm{\uppercase\expandafter{\romannumeral1}})}$ is the polarization of alkali electron spins in Phase \uppercase\expandafter{\romannumeral1}. ${\bf{P}}^{\rm{n}}$ represents the polarization of nuclear spins. Since the nuclear spin polarization remains stable after reaching a steady state, we do not distinguish between Phase \uppercase\expandafter{\romannumeral1}, \uppercase\expandafter{\romannumeral2}, and \uppercase\expandafter{\romannumeral3}, and uniformly represent it using ${\bf{P}}^{\rm{n}}$.
$\gamma_{e/n}$ is the gyromagnetic ratio of electron/nuclear spins. $Q^{(\mathrm{\uppercase\expandafter{\romannumeral1}})}$ is the slowing down factor in Phase \uppercase\expandafter{\romannumeral1}. $\mathbf{B}$ is the external magnetic field. ${\bf{L}}^{(\mathrm{\uppercase\expandafter{\romannumeral1}})}$ represents the light shift induced by the equivalent pump light. $\lambda=8\pi\kappa_0/3$ in a uniformly polarized spherical cell and $\kappa_0$ is the enhancement factor. $M^{\rm{e/n}}_0$ represents the magnetizations of electron or nuclear spins corresponding to full spin polarizations. $\bf{\Omega}$ is the inertial rotation. $R_p^{(\mathrm{\uppercase\expandafter{\romannumeral1}})}$ is the equivalent mean pumping rates of unpolarized atoms of the ground state by the pump light. ${{\bf{S}}_{\bf{p}}}$ is the photon spin of pump light and probe light. $R_{\rm{se}}^{\rm{ne}}$ and $R_{\rm{se}}^{\rm{en}}$ are the spin-exchange rates experienced by alkali and noble gas, respectively. $R_1^{\rm{e{(\mathrm{\uppercase\expandafter{\romannumeral1}})}/n}}$ and $R_2^{\rm{e{(\mathrm{\uppercase\expandafter{\romannumeral1}})}/n}}$ are the longitudinal and transverse relaxation rates of electron or nuclear spins.

Afterward, we use $\mathbf{B^n}=\lambda M^{\rm{{n}}}_0\mathbf{P^n}$ to represent the effective magnetic field produced by noble gas nuclear spins. In Phase \uppercase\expandafter{\romannumeral2}, the evolution of electron spins ${\bf{P}}^{\bf{e}(\mathrm{\uppercase\expandafter{\romannumeral2}})}$ can be expressed by

\begin{align}
\begin{aligned}
\frac{\partial {{\bf{P}}^{\bf{e}(\mathrm{\uppercase\expandafter{\romannumeral2}})}} }{\partial t} =& \frac{{{\gamma _e}}}{{{Q^{(\mathrm{\uppercase\expandafter{\romannumeral2}})}}}}({\bf{B}} + {{\bf{B}}^{\bf{n}}}+ {{\bf{L}}^{(\mathrm{\uppercase\expandafter{\romannumeral2}})}} ) \times {{\bf{P}}^{\bf{e}(\mathrm{\uppercase\expandafter{\romannumeral2}})}} - {\bf{\Omega }} \times {{\bf{P}}^{\bf{e}(\mathrm{\uppercase\expandafter{\romannumeral2}})}} + \frac{{{R_p^{(\mathrm{\uppercase\expandafter{\romannumeral2}})}} {{\bf{S}}_{\bf{p}}} + R_{se}^{ne}{{\bf{P}}^{\bf{n}}}}}{Q^{(\mathrm{\uppercase\expandafter{\romannumeral2}})}} - \frac{{\{ R{{_2^e}^{(\mathrm{\uppercase\expandafter{\romannumeral2}})} },R{{_2^e}^{(\mathrm{\uppercase\expandafter{\romannumeral2}})} },R{{_1^e}^{(\mathrm{\uppercase\expandafter{\romannumeral2}})}}\} }}{Q^{(\mathrm{\uppercase\expandafter{\romannumeral2}})}}{{\bf{P}}^{\bf{e}}}^{(\mathrm{\uppercase\expandafter{\romannumeral2}})} \,,
\end{aligned}
\end{align}
where $Q^{(\mathrm{\uppercase\expandafter{\romannumeral2}})}$ is the slowing down factor in Phase \uppercase\expandafter{\romannumeral2}. ${\bf{L}}^{(\mathrm{\uppercase\expandafter{\romannumeral2}})}$ and $R_p^{(\mathrm{\uppercase\expandafter{\romannumeral2}})}$ are the light shift  and mean pumping rates induced by the real pump light. $R_1^{\rm{e{(\mathrm{\uppercase\expandafter{\romannumeral2}})}}}$ and $R_2^{\rm{e{(\mathrm{\uppercase\expandafter{\romannumeral2}})}}}$ are the longitudinal and transverse relaxation rates of electron spins composed by the real pumping rate $R_p^{(\mathrm{\uppercase\expandafter{\romannumeral2}})}$. According to the evolution in Phase~\uppercase\expandafter{\romannumeral2}, we can obtain the steady state of the electron spins at the end of the pump-on phase %$P^{e(\mathrm{\uppercase\expandafter{\romannumeral2}})}_{x/y/z_\mathrm{end}}$
$P^{e(\mathrm{\uppercase\expandafter{\romannumeral2}})}_{x/y/z}$, which corresponds to the initial state at the beginning of the pump-off phase.

After the pump light is turned off, the longitudinal component of electron spins polarization rapidly depolarizes at a relaxation rate denoted as $R_1^{e(\mathrm{\uppercase\expandafter{\romannumeral3}})}$, which excludes $R_p$. This process results in significant changes in the distribution of the Zeeman sublevels. After a certain period of electron spin decay, the longitudinal component of electron spins polarization approaches zero. Around this point, the variation in the slowing down factor is minimal, and the precession frequency remains relatively stable. We analyze the signals during the period of slow variation in the precession frequency. The alkali-noble gas ensemble operate in a weak magnetic fields and high atomic density, named spin-exchange relaxation-free (SERF), and the transverse relaxation rate $R_2^{e(\mathrm{\uppercase\expandafter{\romannumeral3}})}$ is approximately equal to $R_1^{e(\mathrm{\uppercase\expandafter{\romannumeral3}})}$. We take the moment corresponding to an integer multiple of the precession period during the decay stabilization phase as the initial time. At this phase, the initial value $P^{e(\mathrm{\uppercase\expandafter{\romannumeral3}})}_{x/y/z_\mathrm{0}}$ is given by the product of the decay coefficient $k_e$ and the final state at the end of the pump-on phase $P^{e(\mathrm{\uppercase\expandafter{\romannumeral2}})}_{x/y/z}$. The subsequent evolution of the electron spin can be described by

\begin{align}
\begin{aligned}
\frac{\partial {{\bf{P}}^{\bf{e}(\mathrm{\uppercase\expandafter{\romannumeral3}})}} }{\partial t} =& \frac{{{\gamma _e}}}{{{Q^{(\mathrm{\uppercase\expandafter{\romannumeral3}})}}}}({\bf{B}} + {{\bf{B}}^{\bf{n}}}) \times {{\bf{P}}^{\bf{e}(\mathrm{\uppercase\expandafter{\romannumeral3}})}} - {\bf{\Omega }} \times {{\bf{P}}^{\bf{e}(\mathrm{\uppercase\expandafter{\romannumeral3}})}} + \frac{{ R_{se}^{ne}{{\bf{P}}^{\bf{n}}}}}{Q^{(\mathrm{\uppercase\expandafter{\romannumeral3}})}} - \frac{{\{ R{{_2^e}^{(\mathrm{\uppercase\expandafter{\romannumeral3}})} },R{{_2^e}^{(\mathrm{\uppercase\expandafter{\romannumeral3}})} },R{{_1^e}^{(\mathrm{\uppercase\expandafter{\romannumeral3}})}}\} }}{Q^{(\mathrm{\uppercase\expandafter{\romannumeral3}})}}{{\bf{P}}^{\bf{e}}}^{(\mathrm{\uppercase\expandafter{\romannumeral3}})} \,,
\end{aligned}
\end{align}
where $Q^{(\mathrm{\uppercase\expandafter{\romannumeral3}})}$ is the slowing down factor in Phase \uppercase\expandafter{\romannumeral3}. 

Due to the influence of the effective magnetic field of nuclear spins on the electron spins across three phases, we first derive the nuclear effective magnetic field based on the dynamical equations from Phase \uppercase\expandafter{\romannumeral1} as Eq.~\ref{eqn-Phase1}. In our dynamic polarization configuration, the compensation point $B_c$ is defined as the point where the applied longitudinal magnetic field has a magnitude equal to the sum of the nuclear spin and electron spin effective magnetic field of Phase \uppercase\expandafter{\romannumeral1}, but in the opposite direction (i.e., $B_c=-B_0^n-B_0^{e(\mathrm{\uppercase\expandafter{\romannumeral1}})}$, where $B_0^{n}=\lambda M_0^n P_z^{n}$, $B_0^{e(\mathrm{\uppercase\expandafter{\romannumeral1}})}=\lambda M_0^e P_z^{e(\mathrm{\uppercase\expandafter{\romannumeral1}})}$). When deviating from the compensation point by a small offset (where $\delta B_z=B_z+B_0^n+B_0^{e(\mathrm{\uppercase\expandafter{\romannumeral1}})}$), the $\hat{x}$ and $\hat{y}$ components of nuclear effective magnetic field are
\begin{align}
\begin{aligned}
B_x^n [B]= &\frac{{B_0^n}}{{\delta {B_z} - B_0^n}}{B_x} -  {\frac{{B_0^n}}{{{\gamma _n}\left( {\delta {B_z} - B_0^n} \right)}} } {\Omega _x}-  { \frac{{B_0^n{\gamma _e}\delta {B_z}\left( {{\gamma _n}B{{_0^e}^{(\mathrm{\uppercase\expandafter{\romannumeral1}})}} - \frac{{{\gamma _e}}}{Q}B{{_0^e}^{(\mathrm{\uppercase\expandafter{\romannumeral1}})}}} \right)}}{{\frac{{{\gamma _n}}}{Q}\left( {\delta {B_z} - B_0^n} \right)\left[ {{{\left( {{\gamma _e}\delta {B_z}} \right)}^2} + {{\left( {R{{_2^e}^{(\mathrm{\uppercase\expandafter{\romannumeral1}})}}} \right)}^2}} \right]}}} {\Omega _x} \\
&+ \frac{{B_0^nR{{_2^e}^{(\mathrm{\uppercase\expandafter{\romannumeral1}})}}{\gamma _e}B{{_0^e}^{(\mathrm{\uppercase\expandafter{\romannumeral1}})}}}}{{{\gamma _n}\left( {\delta {B_z} - B_0^n} \right)\left[ {{{\left( {{\gamma _e}\delta {B_z}} \right)}^2} + {{\left( {R{{_2^e}^{(\mathrm{\uppercase\expandafter{\romannumeral1}})}}} \right)}^2}} \right]}}{\Omega _y}\,,\\
B_y^n [B]=& \frac{{B_0^n}}{{\delta {B_z} - B_0^n}}{B_y} -  {\frac{{B_0^n}}{{{\gamma _n}\left( {\delta {B_z} - B_0^n} \right)}} } {\Omega _y} -  \frac{{B_0^n{\gamma _e}\delta {B_z}\left( {{\gamma _n}B_0^e - \frac{{{\gamma _e}}}{Q}B{{_0^e}^{(\mathrm{\uppercase\expandafter{\romannumeral1}})}}} \right)}}{{\frac{{{\gamma _n}}}{Q}\left( {\delta {B_z} - B_0^n} \right)\left[ {{{\left( {{\gamma _e}\delta {B_z}} \right)}^2} + {{\left( {R{{_2^e}^{(\mathrm{\uppercase\expandafter{\romannumeral1}})}}} \right)}^2}} \right]}}{\Omega _y} \\
&- \frac{{B_0^nR_2^e{\gamma _e}B{{_0^e}^{(\mathrm{\uppercase\expandafter{\romannumeral1}})}}}}{{{\gamma _n}\left( {\delta {B_z} - B_0^n} \right)\left[ {{{\left( {{\gamma _e}\delta {B_z}} \right)}^2} + {{\left( {R{{_2^e}^{(\mathrm{\uppercase\expandafter{\romannumeral1}})}}} \right)}^2}} \right]}}{\Omega _x} \,.\label{eq:ComBn}
\end{aligned}
\end{align}

\subsection{Magnetic Field Compensation}
According to Eq.~\ref{eq:ComBn}, it is evident that when $\delta B_z$ equals zero, the nuclear effective magnetic field can fully counteract the external DC magnetic field, thereby realizing the self-compensation.

By substituting this into the dynamical equations for electron spin polarization in Phase $\mathrm{\uppercase\expandafter{\romannumeral2}}$, we derive the steady state solution for electron spin polarization after the pump is activated, with the terms related to the magnetic field expressed as
\begin{align}
    \begin{aligned}
P{_x^{e{(\mathrm{\uppercase\expandafter{\romannumeral2}})}}}[B] = &\frac{{ - R{{_2^e}^{(\mathrm{\uppercase\expandafter{\romannumeral2}})}}{\gamma _e}\frac{{\delta {B_z}}}{{B_0^n}}}}{{{{\left( {R{{_2^e}^{(\mathrm{\uppercase\expandafter{\romannumeral2}})}}} \right)}^2} + {{\left( {{\gamma _e}B{{_0^e}^{(\mathrm{\uppercase\expandafter{\romannumeral1}})}}} \right)}^2}}}P_z^{e{(\mathrm{\uppercase\expandafter{\romannumeral2}})}}{B_y} + \frac{{ - {\gamma _e}^2\frac{{{{\left( {\delta {B_z}} \right)}^2}}}{{B_0^n}} + {\gamma _e}^2B{{_0^e}^{(\mathrm{\uppercase\expandafter{\romannumeral1}})}}\frac{{\delta {B_z}}}{{B_0^n}}}}{{{{\left( {R{{_2^e}^{(\mathrm{\uppercase\expandafter{\romannumeral2}})}}} \right)}^2} + {{\left( {{\gamma _e}B{{_0^e}^{(\mathrm{\uppercase\expandafter{\romannumeral1}})}}} \right)}^2}}}P_z^{e{(\mathrm{\uppercase\expandafter{\romannumeral2}})}}{B_x}\,.\label{eq:Phase2PexSC}
     \end{aligned}
\end{align}

It can be observed that the first derivative of $P{_x^{e{(\mathrm{\uppercase\expandafter{\romannumeral2}})}}}$with respect to $B_y$ exhibits a linear relationship with $\delta B_z$. Likewise, the first derivative of $P{_x^{e{(\mathrm{\uppercase\expandafter{\romannumeral2}})}}}$ with respect to $\delta B_z$ is linearly related to both $B_y$ and $B_x$, while the second derivative of $P{_x^{e{(\mathrm{\uppercase\expandafter{\romannumeral2}})}}}$ with respect to $\delta B_z$ is linearly related to $B_x$. Through the modulation of square waves along the $\hat{y}$ or $\hat{z}$ and by comparing the steady state solutions at high and low levels, we can derive the first derivatives with respect to $B_y$ and $\delta B_z$. Furthermore, another square wave along $\hat{z}$ with varied bias is operated to get the second derivative with respect to $\delta B_z$. By setting the first derivative with respect to $B_y$, the first and second derivative with respect to $\delta B_z$ to zero, three-axis magnetic compensation can be effectively achieved as in the DC mode~\cite{Li2023Magnetic,Gong2025Decoupling}. Moreover, this compensation point fundamentally constitutes a special point for the nuclear spins, which remains essentially stable throughout the evolution of the electron spin. Consequently, self-compensation can be realized across pump-on phase (Phase \uppercase\expandafter{\romannumeral2}) and pump-off phase (Phase \uppercase\expandafter{\romannumeral3}).

\subsection{Response to Inertial Rotation}
Once operating at the compensation point, the response to inertial rotation is more attractive. The nuclear effective magnetic fields for inertial rotation are
\begin{align}
    \begin{aligned}
B_x^n [\Omega]= &\frac{1}{{{\gamma _n}}}{\Omega _x} - \frac{{{\gamma _e}B{{_0^e}^{(\mathrm{\uppercase\expandafter{\romannumeral1}})}}}}{{{\gamma _n}R{{_2^e}^{(\mathrm{\uppercase\expandafter{\romannumeral1}})}}}}{\Omega _y}\,,\\
B_y^n[\Omega] =& \frac{{{\gamma _e}B{{_0^e}^{(\mathrm{\uppercase\expandafter{\romannumeral1}})}}}}{{{\gamma _n}R{{_2^e}^{(\mathrm{\uppercase\expandafter{\romannumeral1}})}}}}{\Omega _x} + \frac{1}{{{\gamma _n}}}{\Omega _y}\,.
     \end{aligned}
\end{align}

Thus, the steady state for $\hat{x}$ and $\hat{y}$ components of electron spins in Phase $\mathrm{\uppercase\expandafter{\romannumeral2}}$ are
\begin{align}
    \begin{aligned}
P{_x^{e{(\mathrm{\uppercase\expandafter{\romannumeral2}})}}}\left[ \Omega  \right] =& \frac{{\frac{{{\gamma _e}P_z^{e{(\mathrm{\uppercase\expandafter{\romannumeral2}})}}}}{{{\gamma _n}}}\left[ {R{{_2^e}^{(\mathrm{\uppercase\expandafter{\romannumeral2}})}} + \frac{{{{\left( {{\gamma _e}B{{_0^e}^{(\mathrm{\uppercase\expandafter{\romannumeral1}})}}} \right)}^2}}}{{R{{_2^e}^{(\mathrm{\uppercase\expandafter{\romannumeral1}})}}}}} \right]{\Omega _y}}}{{{{\left( {R{{_2^e}^{{(\mathrm{\uppercase\expandafter{\romannumeral2}})}}}} \right)}^2} + {{\left[ {{\gamma _e}B{{_0^e}^{(\mathrm{\uppercase\expandafter{\romannumeral1}})}}} \right]}^2}}} + \frac{{\frac{{{\gamma _e}^2B{{_0^e}^{(\mathrm{\uppercase\expandafter{\romannumeral1}})}}P_z^{e{(\mathrm{\uppercase\expandafter{\romannumeral2}})}}}}{{{\gamma _n}}}\left( {\frac{{R{{_2^e}^{(\mathrm{\uppercase\expandafter{\romannumeral2}})}}}}{{R{{_2^e}^{(\mathrm{\uppercase\expandafter{\romannumeral1}})}}}} - 1} \right){\Omega _x}}}{{{{\left( {R{{_2^e}^{(\mathrm{\uppercase\expandafter{\romannumeral2}})}}} \right)}^2} + {{\left[ {{\gamma _e}B{{_0^e}^{(\mathrm{\uppercase\expandafter{\romannumeral1}})}}} \right]}^2}}}\,,\\
P{_y^{e{(\mathrm{\uppercase\expandafter{\romannumeral2}})}}}\left[ \Omega  \right] = &\frac{{ - {\gamma _e}P_z^{e{(\mathrm{\uppercase\expandafter{\romannumeral2}})}}\frac{1}{{{\gamma _n}}}\left[ {R{{_2^e}^{(\mathrm{\uppercase\expandafter{\romannumeral2}})}} + \frac{{{{\left( {{\gamma _e}B{{_0^e}^{(\mathrm{\uppercase\expandafter{\romannumeral1}})}}} \right)}^2}}}{{R{{_2^e}^{(\mathrm{\uppercase\expandafter{\romannumeral1}})}}}}} \right]{\Omega _x}}}{{{{\left( {R{{_2^e}^{(\mathrm{\uppercase\expandafter{\romannumeral2}})}}} \right)}^2} + {{\left[ {{\gamma _e}B{{_0^e}^{(\mathrm{\uppercase\expandafter{\romannumeral1}})}}} \right]}^2}}} + \frac{{\frac{{B{{_0^e}^{(\mathrm{\uppercase\expandafter{\romannumeral1}})}}{\gamma _e}{\gamma _e}}}{{{\gamma _n}}}P_z^{e{(\mathrm{\uppercase\expandafter{\romannumeral2}})}}\left( {\frac{{R{{_2^e}^{(\mathrm{\uppercase\expandafter{\romannumeral2}})}}}}{{R{{_2^e}^{(\mathrm{\uppercase\expandafter{\romannumeral1}})}}}} - 1} \right){\Omega _y}}}{{{{\left( {R{{_2^e}^{(\mathrm{\uppercase\expandafter{\romannumeral2}})}}} \right)}^2} + {{\left[ {{\gamma _e}B{{_0^e}^{(\mathrm{\uppercase\expandafter{\romannumeral1}})}}} \right]}^2}}}\,.
     \end{aligned}
\end{align}

Notably, when $R{_2^{e{(\mathrm{\uppercase\expandafter{\romannumeral1}})}}}=R{_2^{e{(\mathrm{\uppercase\expandafter{\romannumeral2}})}}}$ (with only one Phase), the resulting steady state solution $P{_x^{e{(\mathrm{\uppercase\expandafter{\romannumeral2}})}}}\left[ \Omega  \right] =\gamma_e P_z^{e{(\mathrm{\uppercase\expandafter{\romannumeral2}})}}\Omega_y/\left(R{_2^{e{(\mathrm{\uppercase\expandafter{\romannumeral2}})}}} \gamma_n\right)$ aligns with that of the DC mode~\cite{brown2010new}, thereby confirming the validity of the calculation method. By combining the steady state of Phase $\mathrm{\uppercase\expandafter{\romannumeral2}}$ with the initial state of Phase $\mathrm{\uppercase\expandafter{\romannumeral3}}$ through the decay coefficient $k_e$ we defined before, the expression for the measurement signal $P_x^{e{(\mathrm{\uppercase\expandafter{\romannumeral3}})}}$ during Phase $\mathrm{\uppercase\expandafter{\romannumeral3}}$  is
\begin{align}
    \begin{aligned}
P_x^{e{(\mathrm{\uppercase\expandafter{\romannumeral3}})}}[\Omega]=&\exp \left\{ { - \frac{{R{{_2^e}^{(\mathrm{\uppercase\expandafter{\romannumeral3}})}}t}}{{{Q^{(\mathrm{\uppercase\expandafter{\romannumeral3}})}}}}} \right\}\left[ \begin{array}{l}
\sqrt {{K_{\Omega 1}}^2 + {K_{\Omega 2}}^2} \\
\times \sin \left( {\frac{{{\gamma _e}}}{{{Q^{(\mathrm{\uppercase\expandafter{\romannumeral3}})}}}}B{{_0^e}^{(\mathrm{\uppercase\expandafter{\romannumeral1}})}}t + \arctan \left( {\frac{{{K_{\Omega 2}}}}{{{K_{\Omega 1}}}}} \right)} \right){\Omega _x}\\
 - \sqrt {{K_{\Omega 1}}^2 + {K_{\Omega 2}}^2} \\
 \times \cos \left( {\frac{{{\gamma _e}}}{{{Q^{(\mathrm{\uppercase\expandafter{\romannumeral3}})}}}}B{{_0^e}^{(\mathrm{\uppercase\expandafter{\romannumeral1}})}}t + \arctan \left( {\frac{{{K_{\Omega 2}}}}{{{K_{\Omega 1}}}}} \right)} \right){\Omega _y}
\end{array} \right]\\
 &+ {k_e}P_{{z}}^{e(\mathrm{\uppercase\expandafter{\romannumeral2}})} \exp \left\{ { - \frac{{R{{_1^e}^{(\mathrm{\uppercase\expandafter{\romannumeral3}})}}t}}{{{Q^{(\mathrm{\uppercase\expandafter{\romannumeral3}})}}}}} \right\}\times\frac{{\left(  -  {\frac{1}{{{\gamma _n}}}{\Omega _x} + \frac{{{\gamma _e}B{{_0^e}^{(\mathrm{\uppercase\expandafter{\romannumeral1}})}}}}{{{\gamma _n}R{{_2^e}^{(\mathrm{\uppercase\expandafter{\romannumeral1}})}}}}{\Omega _y}}  \right)}}{{B{{_0^e}^{(\mathrm{\uppercase\expandafter{\romannumeral1}})}}}}\,,
     \end{aligned}
\end{align}
where 
\begin{align}
    \begin{aligned}
{{K_{\Omega 1}}}=&{ - {k_e}P_z^{e(\mathrm{\uppercase\expandafter{\romannumeral2}})}\left[ {\frac{{\frac{{{\gamma _e}}}{{{\gamma _n}}}\left( {R{{_2^e}^{(\mathrm{\uppercase\expandafter{\romannumeral2}})}} + \frac{{{{\left( {{\gamma _e}B{{_0^e}^{(\mathrm{\uppercase\expandafter{\romannumeral1}})}}} \right)}^2}}}{{R{{_2^e}^{(\mathrm{\uppercase\expandafter{\romannumeral1}})}}}}} \right)}}{{{{\left( {R{{_2^e}^{(\mathrm{\uppercase\expandafter{\romannumeral2}})}}} \right)}^2} + {{\left( {{\gamma _e}B{{_0^e}^{(\mathrm{\uppercase\expandafter{\romannumeral1}})}}} \right)}^2}}} - \frac{{{\gamma _e}}}{{{\gamma _n}R{{_2^e}^{(\mathrm{\uppercase\expandafter{\romannumeral1}})}}}}} \right]}\\
{{K_{\Omega 2}}}=&{{k_e}P_z^{e(\mathrm{\uppercase\expandafter{\romannumeral2}})}\left[ {\frac{{\frac{{{\gamma _e}^2B{{_0^e}^{(\mathrm{\uppercase\expandafter{\romannumeral1}})}}}}{{{\gamma _n}}}\left( {\frac{{R{{_2^e}^{(\mathrm{\uppercase\expandafter{\romannumeral2}})}}}}{{R{{_2^e}^{(\mathrm{\uppercase\expandafter{\romannumeral1}})}}}} - 1} \right)}}{{{{\left( {R{{_2^e}^{(\mathrm{\uppercase\expandafter{\romannumeral2}})}}} \right)}^2} + {{\left( {{\gamma _e}B{{_0^e}^{(\mathrm{\uppercase\expandafter{\romannumeral1}})}}} \right)}^2}}} + \frac{1}{{{\gamma _n}B{{_0^e}^{(\mathrm{\uppercase\expandafter{\romannumeral1}})}}}}} \right]}\,.
    \end{aligned}
\end{align}

It can be observed that a specific fitting phase exists, which allows the fitting coefficient of the $\sin$ term to be related to $\Omega_x$, while the $\cos$ term is related to $\Omega_y$. Although it is challenging to ensure in the experiment that the starting point of Phase \uppercase\expandafter{\romannumeral3} coincides with an integer multiple of the precession period, the presence of a time duration $\Delta t$ does not affect the decoupling phenomenon we obtained. According to the relationship that $P_z^{e(\mathrm{\uppercase\expandafter{\romannumeral1}})}=R_p^{(\mathrm{\uppercase\expandafter{\romannumeral1}})}/(R_p^{(\mathrm{\uppercase\expandafter{\romannumeral1}})}+R_\mathrm{sd})=D_r P_z^{e(\mathrm{\uppercase\expandafter{\romannumeral2}})}=D_r R_p^{(\mathrm{\uppercase\expandafter{\romannumeral2}})}/(R_p^{(\mathrm{\uppercase\expandafter{\romannumeral2}})}+R_\mathrm{sd})$, thus, we have $R_p^{(\mathrm{\uppercase\expandafter{\romannumeral1}})}=D_rR_p^{(\mathrm{\uppercase\expandafter{\romannumeral2}})}R_\mathrm{sd}/\left[(1-D_r)R_p^{(\mathrm{\uppercase\expandafter{\romannumeral2}})}+R_\mathrm{sd} \right]$. Suppose that $R_2^{e(\mathrm{\uppercase\expandafter{\romannumeral1}})}\approx R_1^{e(\mathrm{\uppercase\expandafter{\romannumeral1}})}=R_p^{(\mathrm{\uppercase\expandafter{\romannumeral1}})}+R_\mathrm{sd}$, we can simulate the relationship between the scale factor of inertial rotation and the duty ratio as shown in Fig.~\ref{fig:FigKomega}. 
As the duty ratio increases, the scale factor decreases.
%From an intuitive perspective, under the same $\Omega$ input, the transverse nuclear effective magnetic field and the steady state of Phase $\mathrm{\uppercase\expandafter{\romannumeral2}}$ exhibit minimal changes. As the duty ratio increases, the longitudinal nuclear effective magnetic field increases. Consequently, the angle between the initial electron polarization of Phase $\mathrm{\uppercase\expandafter{\romannumeral3}}$ and the magnetic field axis around which the precession occurs decreases, leading to a smaller scale factor.

\begin{figure}[htbp]
\centering   
\includegraphics[width=0.3\linewidth]{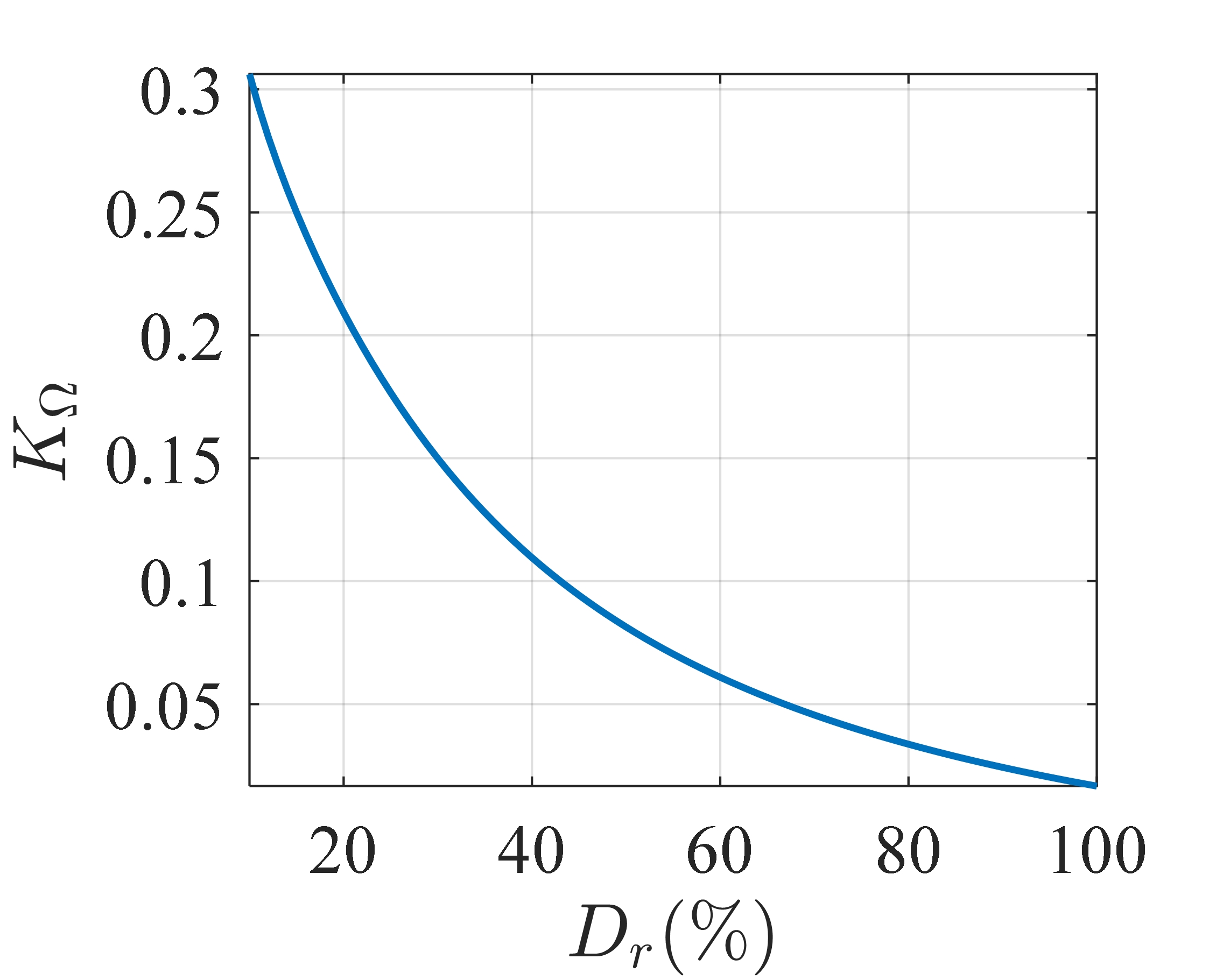}%
\caption{Simulation for the relationship between the scale factor ($K_{\Omega}=\sqrt{K_{\Omega1}^2+K_{\Omega2}^2}$) of inertial rotation and the duty ratio in Phase \uppercase\expandafter{\romannumeral3}. It can be seen that as the duty ratio increases, the scale factor for inertial rotation decreases.}
\label{fig:FigKomega}
\end{figure}

\subsection{Suppression to Low-Frequency Magnetic Field}
When the oscillating transverse magnetic field ${B_x}\cos \left( {\omega t} \right)\hat{x}+{B_y}\cos \left( {\omega t} \right)\hat{y}$ with the frequency $\omega$ and the amplitude $B_x$ or $B_y$ occurs, the $\hat{x}$ and $\hat{y}$ components of nuclear effective magnetic field are
\begin{align}
\begin{aligned}
B_x^n(\omega) = & - {B_x}\cos \left( {\omega t} \right) - \frac{{{\gamma _e}B_0^{e(\mathrm{\uppercase\expandafter{\romannumeral1}})}R_2^n}}{{R_2^{e(\mathrm{\uppercase\expandafter{\romannumeral1}})}{\gamma _n}B_0^n}}{B_x}\cos \left( {\omega t} \right) - \frac{{\omega {\gamma _e}B_0^{e(\mathrm{\uppercase\expandafter{\romannumeral1}})}}}{{R_2^{e(\mathrm{\uppercase\expandafter{\romannumeral1}})}{\gamma _n}B_0^n}}{B_x}\sin \left( {\omega t} \right) \\
& + \frac{{R_2^n{B_y}\cos \left( {\omega t} \right)}}{{{\gamma _n}B_0^n}} - \frac{{\omega {B_y}\sin \left( {\omega t} \right)}}{{{\gamma _n}B_0^n}}\,,\\
B_y^n (\omega)=  &- {B_y}\cos \left( {\omega t} \right) - \frac{{R_2^n{B_x}\cos \left( {\omega t} \right)}}{{{\gamma _n}B_0^n}} + \frac{{\omega {B_x}\sin \left( {\omega t} \right)}}{{{\gamma _n}B_0^n}}\\
& - \frac{{{\gamma _e}B_0^{e(\mathrm{\uppercase\expandafter{\romannumeral1}})}R_2^n}}{{R_2^{e(\mathrm{\uppercase\expandafter{\romannumeral1}})}{\gamma _n}B_0^n}}{B_y}\cos \left( {\omega t} \right) - \frac{{\omega {\gamma _e}B_0^{e(\mathrm{\uppercase\expandafter{\romannumeral1}})}}}{{R_2^{e(\mathrm{\uppercase\expandafter{\romannumeral1}})}{\gamma _n}B_0^n}}{B_y}\sin \left( {\omega t} \right)\,.\label{eq:BnBxBycos}
\end{aligned}
\end{align}

In each expression, the first term serves to cancel the external input magnetic field, while the remaining terms are undesired interference terms. We can also derive the low-frequency response of the steady state in Phase $\mathrm{\uppercase\expandafter{\romannumeral2}}$ as
\begin{align}
\begin{aligned}
P{_x^{e(\mathrm{\uppercase\expandafter{\romannumeral2}})}}  (\omega)=& {\gamma _e}P_z^{e(\mathrm{\uppercase\expandafter{\romannumeral2}})}\times\left[\begin{array}{l}
 - \frac{{R_2^n}}{{{\gamma _n}B_0^n}}\left( {R{{_2^e}^{(\mathrm{\uppercase\expandafter{\romannumeral2}})}} - \frac{{{{\left( {{\gamma _e}B{{_0^e}^{(\mathrm{\uppercase\expandafter{\romannumeral1}})}}} \right)}^2}}}{{R{{_2^e}^{(\mathrm{\uppercase\expandafter{\romannumeral1}})}}}}} \right){B_x}\cos \left( {\omega t} \right)\\
 + \frac{\omega }{{{\gamma _n}B_0^n}}\left( {R{{_2^e}^{(\mathrm{\uppercase\expandafter{\romannumeral2}})}} + \frac{{{{\left( {{\gamma _e}B{{_0^e}^{(\mathrm{\uppercase\expandafter{\romannumeral1}})}}} \right)}^2}}}{{R{{_2^e}^{(\mathrm{\uppercase\expandafter{\romannumeral1}})}}}}} \right){B_x}\sin \left( {\omega t} \right)\\
 - \frac{{{\gamma _e}B{{_0^e}^{(\mathrm{\uppercase\expandafter{\romannumeral1}})}}R_2^n}}{{{\gamma _n}B_0^n}}\left( {\frac{{R{{_2^e}^{(\mathrm{\uppercase\expandafter{\romannumeral2}})}}}}{{R{{_2^e}^{(\mathrm{\uppercase\expandafter{\romannumeral1}})}}}} + 1} \right){B_y}\cos \left( {\omega t} \right)\\
 - \frac{{\omega {\gamma _e}B{{_0^e}^{(\mathrm{\uppercase\expandafter{\romannumeral1}})}}}}{{{\gamma _n}B_0^n}}\left( {\frac{{R{{_2^e}^{(\mathrm{\uppercase\expandafter{\romannumeral2}})}}}}{{R{{_2^e}^{(\mathrm{\uppercase\expandafter{\romannumeral1}})}}}} - 1} \right){B_y}\sin \left( {\omega t} \right)
\end{array}\right] \times\frac{1}{{{{\left( {R{{_2^e}^{(\mathrm{\uppercase\expandafter{\romannumeral2}})}}} \right)}^2} + {{\left( {{\gamma _e}B{{_0^e}^{(\mathrm{\uppercase\expandafter{\romannumeral1}})}}} \right)}^2}}}\,.
\end{aligned}
\end{align}

The response of $P{_y^{e(\mathrm{\uppercase\expandafter{\romannumeral2}})}}$ is similar with $P{_x^{e(\mathrm{\uppercase\expandafter{\romannumeral2}})}}$. In $P{_y^{e(\mathrm{\uppercase\expandafter{\romannumeral2}})}}$, the response coefficient for the magnetic field along $\hat{y}$ matches that for the $\hat{x}$ field in $P{_x^{e(\mathrm{\uppercase\expandafter{\romannumeral2}})}}$, whereas the response coefficient for the $\hat{x}$ field in $P{_y^{e(\mathrm{\uppercase\expandafter{\romannumeral2}})}}$ differs from that for the $\hat{y}$ field in $P{_x^{e(\mathrm{\uppercase\expandafter{\romannumeral2}})}}$ only by a sign. The simulated response of low-frequency magnetic field is shown in Fig.~\ref{fig:FigBxBycosSimu}(a). It is evident that for both $\hat{x}$ and $\hat{y}$ magnetic field, higher duty ratio demonstrates superior suppression effect, resulting in smaller response. Furthermore, under high duty ratio (higher $B_0^n$, $B_0^{e(\mathrm{\uppercase\expandafter{\romannumeral1}})}$ and $R_2^{e(\mathrm{\uppercase\expandafter{\romannumeral1}})}$), according to Eq.~\ref{eq:BnBxBycos}, the nuclear effective magnetic field has smaller amplitude. Consequently, in Phase $\mathrm{\uppercase\expandafter{\romannumeral3}}$ , the $\hat{x}$ and $\hat{y}$ projections of the distance between the initial electron polarization (which depends on $(P{_x^{e(\mathrm{\uppercase\expandafter{\romannumeral2}})}},P{_y^{e(\mathrm{\uppercase\expandafter{\romannumeral2}})}},P{_z^{e(\mathrm{\uppercase\expandafter{\romannumeral2}})}})$) and its precession axis (which depends on the direction of $(B_x+B_x^n,B_y+B_y^n,-B{{_0^e}^{(\mathrm{\uppercase\expandafter{\romannumeral1}})}})$) decreases as the duty ratio increases when $B_x\cos(\omega t)$ exists, as shown in Fig.~\ref{fig:FigBxBycosSimu}(b). As for the $B_y\cos(\omega t)$, The amplitudes of the two components are exchanged. Thus, increasing the duty ratio results in a smaller response, in other words, a higher suppression.

\begin{figure}[htbp]
\centering   
\includegraphics[width=0.5\linewidth]{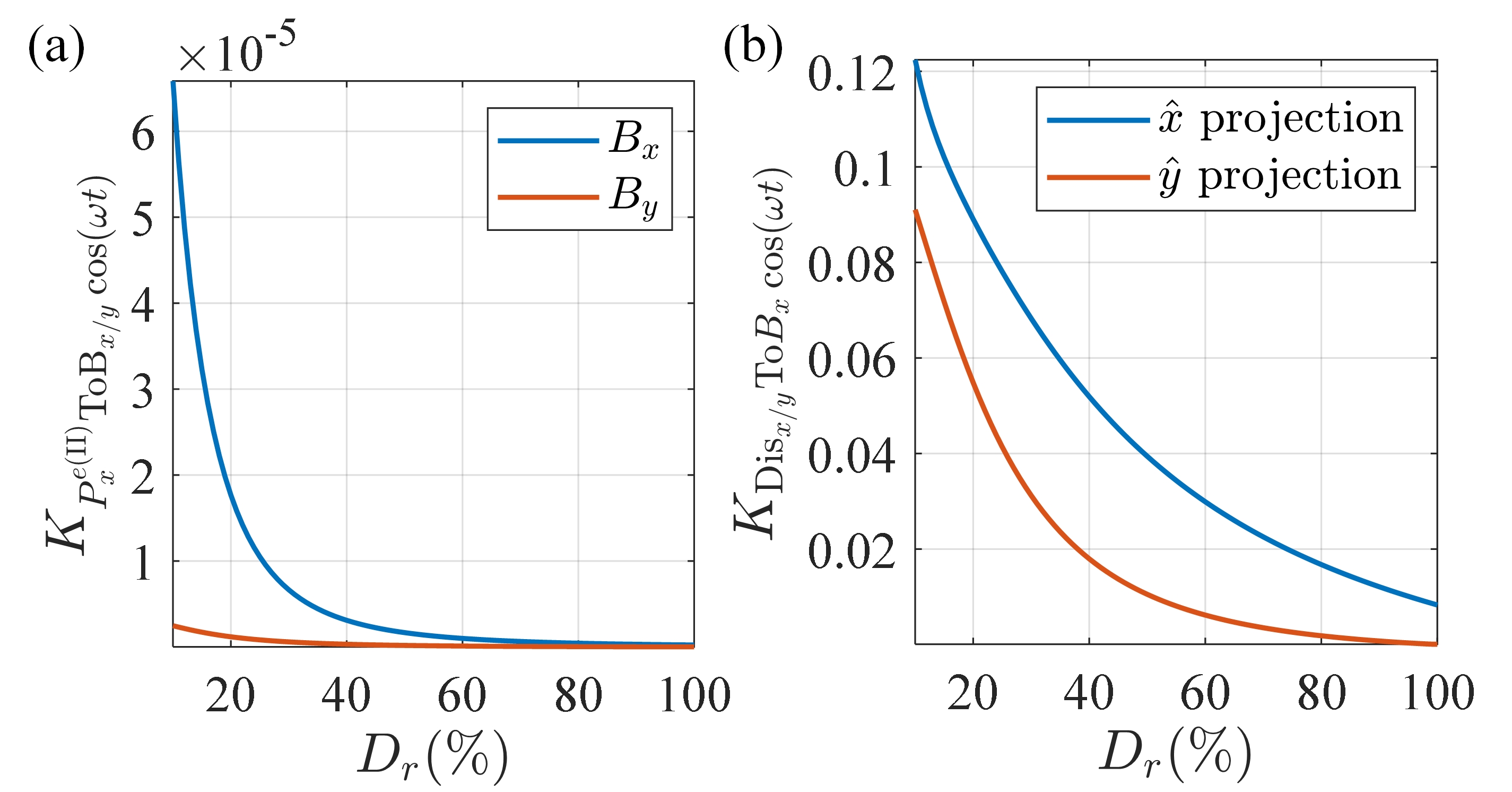}%
\caption{Simulation for the relationship between the scale factor of low-frequency magnetic field and the duty ratio. (a) In Phase \uppercase\expandafter{\romannumeral2}, as the duty ratio increases, the steady state response $P{_x^{e(\mathrm{\uppercase\expandafter{\romannumeral2}})}}$ to the low-frequency magnetic field decreases, thereby strengthening the suppression ability. (b) The $\hat{x}$ and $\hat{y}$ projection of the distance between the steady state of Phase \uppercase\expandafter{\romannumeral2} and the precession axis when $B_x\cos(\omega t)$ exists, which is related to the damped oscillation amplitude in Phase \uppercase\expandafter{\romannumeral3}, decreases as the duty ratio increases. }
\label{fig:FigBxBycosSimu}
\end{figure}

\section*{Resource availability}

\subsection*{Lead contact}

Further information and requests for resources should be directed to the lead contact Kai Wei (weikai@buaa.edu.cn).

\subsection*{Materials availability}

This paper did not generate new materials.

\subsection*{Data and code availability}
All data reported in this paper will be shared by the lead contact upon request.

This paper does not report original code.

Any additional information required to reanalyze the data reported in this paper is available from the lead contact upon request.

\section*{Acknowledgements}
This work is supported in part by the Innovation Program for Quantum Science and Technology under Grant 2021ZD0300401, and in part by the National Natural Science Foundation of China (NSFC) under Grant 62203030 and Grant 61925301 for Distinguished Young Scholars, and in part by the Fundamental Research Funds for the Central Universities.

\section*{Author contributions}
X.H., K.W., and Y.R. proposed this study. X.H., D.G., S.Z., and J.Z. 
performed the experiment and analyzed the data. X.H., K.W., and W.Q. wrote the manuscript.

\section*{Declaration of interests}
The authors declare no competing interests.

\section*{Inclusion and diversity}
We support inclusive, diverse, and equitable conduct of research.

\end{widetext}

\end{document}